\documentclass[a4paper,11pt]{article}
\usepackage[utf8]{inputenc}

\usepackage{jcappub}

\usepackage{amsmath}
\usepackage{amssymb}
\usepackage{physics}
\usepackage{subcaption}
\usepackage{graphicx}
\usepackage{tabu}
\usepackage{makecell}
\usepackage{hyperref}
\usepackage{cleveref}
\usepackage{float}
\usepackage{enumitem}

	\newcommand{\keV}{\,\mathrm{keV}}
	\newcommand{\MeV}{\,\mathrm{MeV}}
	\newcommand{\GeV}{\,\mathrm{GeV}}
	\newcommand{\TeV}{\,\mathrm{TeV}}
	\newcommand{\Lya}{Lyman-$\alpha$ forest}

\title{Probing non-thermal light DM with structure formation and $N_\mathrm{eff}$}
\author[a]{Sven Baumholzer}
\author[a]{Pedro Schwaller}

\affiliation[a]{\textit{$\text{PRISMA}^+$ Cluster of Excellence \& Mainz Institute for Theoretical Physics,
	Johannes \\Gutenberg-Universität Mainz, Staudingerweg 7, 55099 Mainz, Germany}}

\emailAdd{baumholz@uni-mainz.de}
\emailAdd{pedro.schwaller@uni-mainz.de}

\begin{document}

\abstract{
In many models of dark matter (DM), several production mechanisms contribute to its final abundance, often leading to a non-thermal momentum distribution. This makes it more difficult to assess whether such a model is consistent with structure formation observations. 
We simulate the matter power spectrum for DM scenarios characterized by at least two temperatures and derive the suppression of structures at small scales and the expected number of Milky Way dwarf galaxies from it. This, together with the known bound on the number of relativistic 
particle species, $N_\mathrm{eff}$, allows us to obtain constraints on the parameter space of non-thermally produced DM. 
We propose a simple parametrization for non-thermal DM distributions and present a fitting procedure that can be used to adapt our results to other models. 
}

\begin{flushright}
  MITP-21-069
\end{flushright}

\maketitle
\clearpage
\section{Introduction}
Although the WIMP paradigm is strictly speaking not ruled out yet, it is definitely under pressure from current direct detection results \cite{Aprile:2018dbl} and hence the last several years have seen a rise in interest of other avenues 
to explain the DM puzzle. Specifically, the focus has shifted to DM candidates of lighter masses. While these particles allow for potential new experimental probes, they may have an impact on the cosmology of the early 
universe as well. In fact, if DM is light enough such that it is still relativistic at sufficiently late times it has a non-vanishing free streaming length and will change the formation and the properties of galaxies compared to the standard $\Lambda$CDM paradigm 
\cite{Bode:2000gq} below this length scale. 
\\
Initially, the effect of such ``hot" DM has been studied in the context of SM neutrinos as a DM candidate, which turned out to feature a too large free streaming length, effectively erasing structures at far too large scales. Consequently, 
attention has shifted to DM models which can be considered to be ``warm" instead. Interestingly, $N$-body simulations of structure formation in the $\Lambda$CDM regime revealed small scale structures which are in tension with observations
(for an overview, see \cite{Bullock:2017xww}). These simulations predict too large numbers of accompanying galaxies of the Milky Way (MW) \cite{Moore:1999nt,Klypin:1999uc}, which is called the 
\textit{missing satellite problem}. Further, the shape of galactic cores do not match with observations \cite{de_Blok_2010} (the \textit{cusp versus core problem}) and lastly, the \textit{too big too fail problem} \cite{Boylan_Kolchin_2011,Boylan_Kolchin_2012} addresses a mismatch in the dynamics of the brightest MW satellites.
It is still under debate if these issues can be alleviated in the $\Lambda$CDM paradigm by including baryonic feedback in the simulations (see for instance \cite{Zolotov_2012,Dutton:2015nvy,Lovell:2016nkp}).
\\
On the other hand, going beyond the $\Lambda$CDM paradigm, aforementioned tensions can be cured by invoking ``warm" DM models which alter the small scale structures but agree with cold DM (CDM) at large scales.  
Of particular interest with respect to above mentioned questions are DM mixtures of cold and warm DM species. Often, the cold component is considered to be of standard $\Lambda$CDM origin, while the warm component is made of something else: popular extensions involve sterile neutrinos \cite{Boyarsky:2008xj, Schneider:2014rda,Merle:2015vzu}, ultralight particles \cite{Kobayashi:2017jcf,Hui:2016ltb}, axions 
\cite{Marsh:2013ywa}, fuzzy DM \cite{Hu:2000ke,Schwabe:2020eac} or non-cold thermal relics \cite{Diamanti:2017xfo,Anderhalden_2012}. Some of the first models invoked neutrinos as a hot DM candidate mixed with a cold component (see for instance 
\cite{Davis:1992ui,Klypin:1992sf,Klypin:1994iu,Ma:1994ub,Dodelson:1995es}). 
\\
In this article, we are mainly concerned with scenarios with only a single DM species, which however features a non-thermal momentum distribution, for example due to different production mechanisms contributing to its relic abundance. In freeze-in scenarios, the DM momentum distribution is a priori unknown, and depends on the exact production mode. If DM is produced from the decays of a heavy parent particle, the crucial question is whether at the time of decay the parent is in thermal equilibrium~\cite{Hall:2009bx} or itself frozen out, as in the Super-WIMP mechanism~\cite{Covi:1999ty,Feng:2003uy}. If more than one parent particle makes an appreciable contribution to the final DM abundance, a highly non-thermal spectrum featuring several characteristic peaks is obtained.
\\
Such scenarios of DM with a non-thermally produced warm admixture have been considered for instance in \cite{Hooper:2011aj, Heeck:2017xbu, Garny:2018ali, Feng:2003uy,Baumholzer:2018sfb,Baumholzer:2019twf,Decant:2021mhj}. Precisely, we will 
have a subset of DM characterized by a higher temperature such that it can be 
considered ``warm" or even ``hot". 
The aim is to quantify our results such that they can be mapped onto a wide variety of models which might feature other ways to produce DM at different times. A similar idea was done in \cite{Parimbelli:2021mtp}, where the authors trained an emulator using the matter power spectrum of a mixed cold and warm DM model setup. Finally, the authors of \cite{Dienes:2020bmn, Dienes:2021itb} discuss the effect of non-minimal dark sector momentum distribution functions on cosmology.
\\
We introduce a model-independent parametrization of the DM momentum distribution function and use current limits on observables related to the matter power spectrum to set bounds on the allowed parameter space. We dub our framework the \textit{two temperature dark matter} (2TDM). 
The paper is organized as follows: in \cref{sec:model_setup} we discuss how we parametrize the 2TDM and explain the assumptions we made when modeling the production mechanism. \Cref{sec:constraints_on_model_parameter} consists of a discussion of 
constraints on the parameter space derived from cosmological observations. In \cref{sec:going_beyond_thermal} we discuss some modifications of the simplified description. Results are shown in \cref{sec:fit_results} and an 
application of the matching is given in \cref{sec:application_toy_models}. Finally, we summarize our results in \cref{sec:summary}. 
\section{Model setup}
\label{sec:model_setup}
\begin{figure}[ht!]
	\centering
	\includegraphics[width=0.6\textwidth]{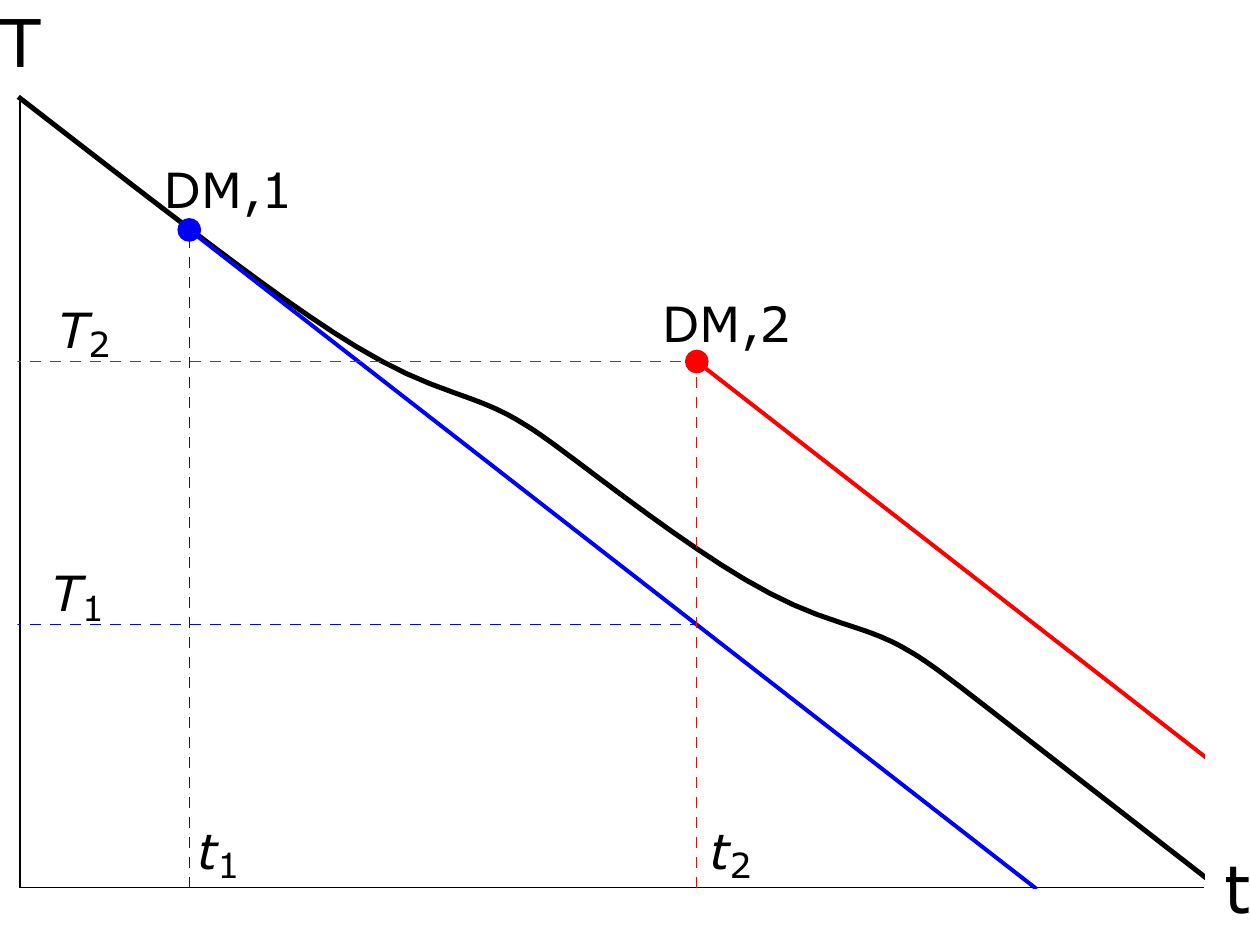}
	\caption{Schematic representation of our setup: given in black is the background temperature $T_\gamma$ of the universe, while the blue and red curve represents the time evolution of the temperature of the first and second DM set, respectively. The first DM subset is produced at time $t_1$ with a temperature similar to $T_\gamma$. After a while, the second DM subset is produced at $t_2$ but it has a higher temperature $T_2$ compared to $T_1$. The bumps in the solid black line mimics entropy dilution due to particle freeze-out in the SM thermal bath.}
	\label{fig:temperature_scheme}
\end{figure}

We consider a model where the observed DM density is explained via one particle species which is produced in two different production modes at separate times. Therefore, an intrinsic temperature can be assigned to both 
production possibilities. As such the final DM abundance is made up by two shares of a single particle species each with its own temperature. For convenience we are going to refer to the earlier produced part as the \textit{first subset}, while the other one is the \textit{second subset}. A schematic representation of this is shown in \cref{fig:temperature_scheme}. Crucially, the second DM subset whose production happens at a later time $t_2$ features a significantly larger 
temperature than the first DM set which is produced earlier on at time $t_1$, that means we take that $T_2>T_1$.\footnote{This does not have to be necessarily true for all cases. Depending on the properties of the parent species, one can have model setups, where the temperature of the latter produced DM subset is smaller compared to the temperature of the DM produced at earlier times. However, in that case one can change the naming $1\leftrightarrow2$ for both DM subsets. }
\\
A scenario with these assumptions has important consequences for the behavior of DM. In fact, while the first subset might be cold, the second one will be warm or even hot DM. The question which arises is two-fold: how large can the temperature of the second subset be, and how much of it can be produced?
To assess these questions, we will study the impact of the 2TDM on the structure formation at galactic scales. To be precise, as will be explained in \cref{sec:limits_from_structure_formation}, we are adopting observations of the Lyman-$\alpha$ forest and the observed number of MW subhalos to constrain the 2TDM.
A visualization of the parameters we are constraining is given in \cref{fig:DM_spectrum}, where we show an example of the DM momentum distribution functions $x^2\,f(x)$, where  $x=p/T$, which is a characteristic quantity for a given DM model. The spectrum features two distinctive peaks, which can be characterized by two quantities: first, the position of their peaks which is related to the respective DM temperature and second, their contribution to the total DM density $\Omega_\text{DM} h^2$ which is set by the area under the respective curve. As can be seen, in this example the first share is the dominant contribution which contributes a part $A_1$ to the total DM density (blue shaded region), while the hotter DM part contributes $A_2$ (red shaded region).
\subsection{Parametrization}
\label{subsec:decay_parammetrization}
\begin{figure}[t]
	\centering
	\includegraphics[width=0.8\textwidth]{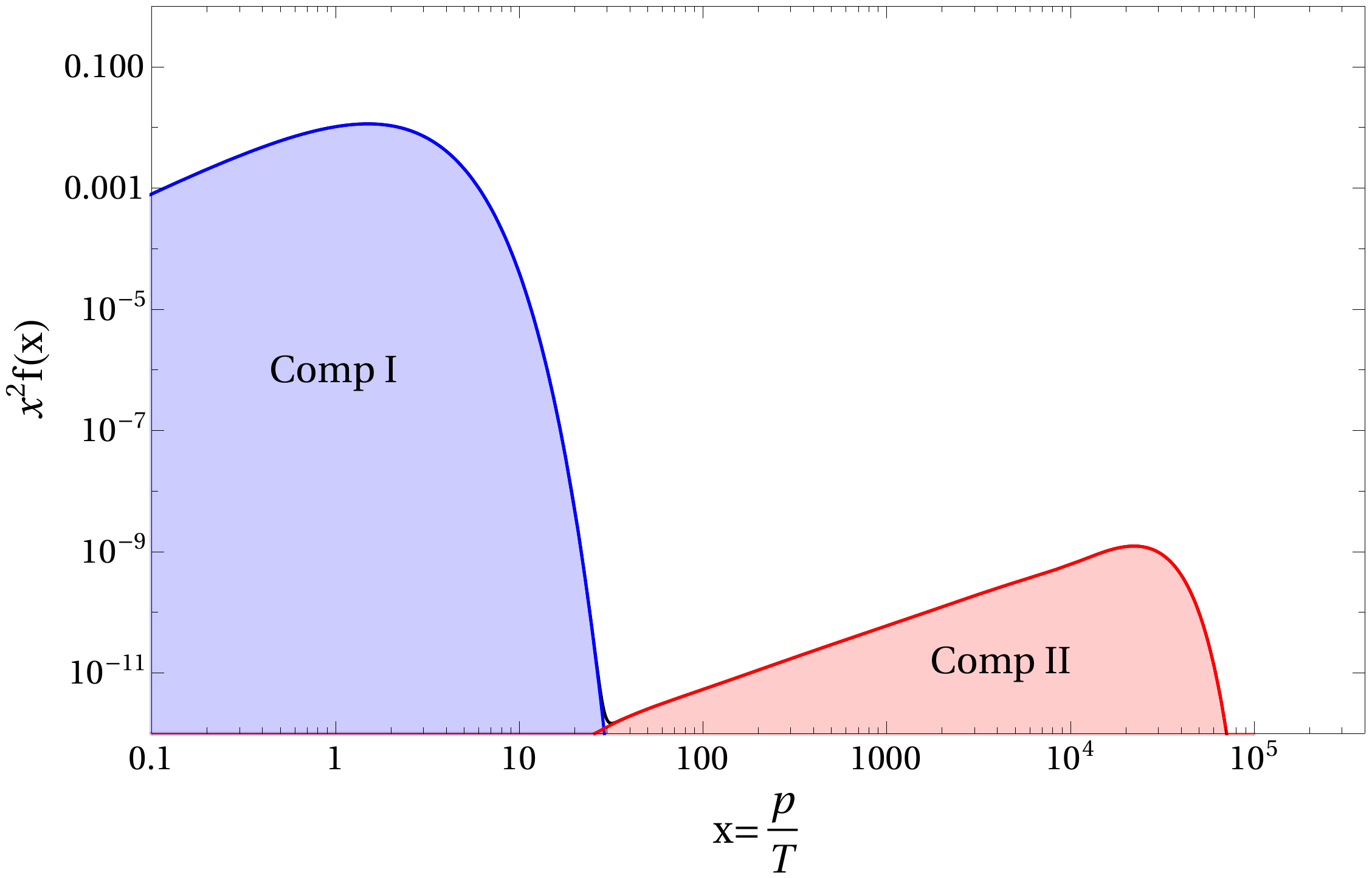}
	\caption{Example $x^2\,f(x)$ spectrum for a specific 2TDM scenario with two observable peaks calculated for times when DM production is finished. The blue area corresponds to a dominant subset whereas the red area refers to a subdominant DM share but with higher temperature with respect to the first set. }
	\label{fig:DM_spectrum}
\end{figure}

The momentum distribution shown in \cref{fig:DM_spectrum} corresponds to DM freeze-in production via decays of thermalized (blue curve) and non-thermal particles (red curve). Following this, we are going to assume that DM in the 2TDM is produced via decays of heavy parent particles irrespective whether they were previously thermalized or not.
\\
Specifically, we are interested in the following two parameters:
\begin{align*}
\text{Temperature ratio:}& \quad \xi \equiv \frac{T_2}{T_1} >1,\\
\text{Abundance}: &\quad A_2\in\left(0,1\right]. 
\end{align*}
The abundance $A_1$ of the first DM subset is fixed by the requirement $A_1 = 1 - A_2$ such that the DM relic abundance is achieved.
\\
It is important to note that we assume the temperature between both DM subsets to stay constant once produced during the subsequent time evolution of the universe. In principle, one could have entropy dilution in the dark sector similar to the SM thermal bath, but in the following we want to neglect such dilution effects for the dark sector.\\

\subsection{Modeling the decays}
\label{subsec:analytical_modeling}
In the following, we model the production of the individual DM subsets by decays of parent particles $P$, which decay into DM particles $X$ via $P \to X\,X$
and whose mass is $m_P$. More details on the DM production via decaying scalars can be found for instance in \cite{Petraki:2007gq,Merle:2013wta, Merle_2015, Konig:2016dzg}. In the following we want to focus solely on analytical expressions derived in \cite{Merle_2015} which corresponds to the following types of production:
\begin{itemize}
	\item The parent particle decays while it is still in thermal equilibrium.
	\item For sufficiently small decay widths, the scalar particle freeze-out before it decays into DM.
	\item If the scalar is only weakly coupled to the SM, it never thermalizes before or during its decays, but freezes in instead.
\end{itemize}
Solving the Boltzmann equation for the DM momentum distribution yields \cite{Merle_2015}
\begin{equation}
	f(x,r)=2\frac{C_\Gamma}{g_*(T_\mathrm{prod})}\,\int\limits_0^r\dd s\, \frac{s^2}{x^2}\int\limits_{\tilde{x}_\text{min}}^\infty\dd \tilde{x} \frac{\tilde{x}}{\sqrt{\tilde{x}^2 + s^2}}f_P(\tilde{x},s).
	\label{eq:master_equation}
\end{equation}
Here, $C_\Gamma$ denotes an effective DM production rate via decays of $P$. As such it is related to the decay width\footnote{We assume that there is only one decay channel for $P$ and so $\Gamma$ is a total decay width.} $\Gamma$ of $P$ by $\displaystyle C_\Gamma =M_0\,\Gamma/m_P^2$, where $M_0 =7.35\cdot 10^{18}\GeV$.  The factor of two in front of the expression has to be dropped if one looks at processes $P \to X\,S$ instead. Although it is implicitly assumed for this derivation that the number of effective entropic degrees of freedom, $g_*(T)$, are not changing at all, a dimensionless time variable $r=m_P/T$ is introduced to track changes in $g_*(T)$ during DM production; for a general discussion we refer to \cref{sec:modifications_variable_dof}.\\
In the following, we present the result for parent particles decaying while maintaining thermal equilibrium, whereas the other two cases are given in \cref{subsec:decay_freeze_out_scalar,subsec:decay_non_thermal_scalar} in the appendix. Assuming a Maxwell-Boltzmann (MB) distribution\footnote{This assumption is crucial to derive analytic results, while the inclusion of a Bose-Einstein or Fermi-Dirac distribution only marginally changes the result. As such our discussion holds for all types of parent particles.} for the parent particle,\\ $\displaystyle f_P(x,r) =\text{exp}\left(- \sqrt{r^2+x^2} \right)$, one can derive the following expression for the DM momentum distribution \cite{Merle_2015}:
\begin{equation}
	f(x,r) = 8C_\Gamma e^{-x}\left(  \frac{r}{2x} e^{-r^2/4x} + \frac{1}{2} \sqrt{\frac{\pi}{x}} \text{Erf}\left[ \frac{r}{\sqrt{4x}} \right] \right).
	\label{eq:equilibrium_decays}
\end{equation}
In the limit $r\to \infty$, \cref{eq:equilibrium_decays} reduces to
\begin{equation}
	f(x,\infty) \equiv f(x)= 4C_\Gamma e^{-x}\sqrt{\frac{\pi}{x}}.
	\label{eq:equ_momentum_distr}
\end{equation}
From this expression we can deduce today's DM density
\begin{equation}
\Omega_\text{DM}h^2 = \frac{s_0 m_\text{DM}}{\rho_\text{crit}/h^2}\left(\frac{45g}{4\pi^4g_*(T_\text{prod})}\right) \int\limits_{0}^{\infty}\dd x \, x^2 f(x,\infty),
\label{eq:omega_density}
\end{equation}
where the entropy density is $s_0= 2891.2\,\text{cm}^{-3}$ \cite{Zyla:2020zbs} and the critical density is given by $\rho_\text{crit}=1.054\cdot 10^{-2} \MeV\,\text{cm}^{-3}\,h^2$ \cite{Zyla:2020zbs}. Inserting \cref{eq:equ_momentum_distr} into \cref{eq:omega_density} one can derive
\begin{equation}
	\Omega_\text{DM}h^2 = C_\Gamma K \frac{m_\text{DM}}{\text{GeV}}\frac{g}{g_*(T_\text{prod})},
	\label{eq:omega_density_simplified}
\end{equation}
where $K \simeq 3\cdot10^{8}$ and $g$ denotes the internal degrees of freedom of the DM species. 
It may seem counterintuitive at first that the DM density is directly proportional to the decay width of $P$, but under this approximation the parents do not deviate from a MB distribution and hence DM can only be efficiently produced before the parent particles experience a Boltzmann suppression at temperatures $T\approx m_P$. The corresponding abundances for the other two production processes do not depend on the size of $C_\Gamma$.\\
The above formulas can be straightforwardly generalized to a DM scenario composed of $i$ subsets, each with a temperature $T_i$. In that case we are defining a reference temperature $T_1$ which we take to be given by the lowest temperature of the DM species and so the final momentum distribution for the DM, $f_\text{DM}(x_i)$, is given by the following sum:
\begin{equation}
f_\text{DM}(x_1) = \sum_i f_i(x_i) = 4\sum_i C_{\Gamma,i}\sqrt{\frac{\pi}{x_i}} e^{-x_i} = 4\frac{\pi}{\sqrt{x_1 T_1}} \sum_i C_{\Gamma,i} \sqrt{T_i} e^{-x_1 T_1/T_i}.
\label{eq:shifted_MB_distr}
\end{equation}
Here each subspecies has its own decay width $C_{\Gamma,i}$ and temperature $x_i=p/T_i$. For the 2TDM we find that
\begin{align}
\label{eq:DM_momentum_distr_func}
f_\text{DM}(x_1) & =   4\sqrt{ \frac{\pi}{x_1}} \left( C_{\Gamma,1} e^{-x_1} + C_{\Gamma,2} \sqrt{\xi} e^{-x_1 / \xi} \right), \\
\Omega_\mathrm{DM}h^2 & =  K \frac{m_\text{DM}}{\text{GeV}}g \left( \frac{C_{\Gamma,1}}{g_*(T_\text{prod,1})}  + \frac{C_{\Gamma,2}}{g_*(T_\text{prod,2})} \xi^3 \right) \notag\\
 &\equiv \Omega_\mathrm{DM}h^2 \left( A_1 + A_2 \right).
\label{eq:Omega_h_combined}
\end{align}
Thus, \cref{eq:Omega_h_combined} allows to relate $A_1$ and $A_2$ to the decay width $C_{\Gamma,i}$ and demanding that the DM relic abundance is generated constrains $C_{\Gamma,1},\;C_{\Gamma,2}$ and $\xi$.
However, above expressions are only valid if the respective parent particles are in thermal equilibrium during their decay into DM.
\\
Next we want to match our prescription onto DM produced by late decaying particles by defining a relation between the production rate $C_\Gamma$ and the temperature ratio of the 2TDM. We start by comparing the averaged momentum $\langle x_2 \rangle$ for the late time produced particles against $\langle x_1 \rangle$ stemming from DM production of thermalized parent particles. The respective averaged momenta are given by
\begin{equation}
	\langle x_2 \rangle = \frac{\int_0^\infty \dd x_1 \, x_1^3 f(x_1/\xi)}{\int_0^\infty \dd x_1 \, x_1^2 f(x_1/\xi)} = 2.5\,\xi = \xi \langle x_1 \rangle\,,
	\label{eq:shifted_avg_value}
\end{equation}
where the averaged value $\langle x_2\rangle$ is shifted by a factor $\xi$ compared to $\langle x_1\rangle=2.5$. This is important for matching the temperature ratio $\xi$ to a specific decay width of long-lived parent particles.\\
For this matching we assume the parent particles to be at rest when decaying and producing the DM. Further we assume an instantaneous decay at time $\tau = t_{1/2} = \log 2 /\Gamma$ set by the half-time of the parent particle, which can be converted into a temperature by the relation
\begin{align}
		\frac{T}{\text{MeV}} = & \,1.55\, g_*(T)^{-1/4} \left(\frac{\text{sec}}{\tau}\right)^{1/2} \notag\\
		\simeq & \,\,8.5\cdot 10^5\,\frac{m_P}{\text{TeV}} \,g_*(T)^{-1/4}\sqrt{C_\Gamma}\,.
		\label{eq:temp_decay_relation}
\end{align}
In the last step we used that $\Gamma = m_P^2\,C_\Gamma/M_0$. This temperature has to be compared to the energy of the DM which is roughly set by $E \approx p = m_P/2$. 
Finally, we can equate $p/T$ and \cref{eq:shifted_avg_value} to derive a relation between our model parameter $\xi$ and the DM production rate $C_\Gamma$
\begin{equation}
\xi = \frac{m_P}{5T(C_\Gamma)} \quad \Longrightarrow \quad \xi \simeq 0.24\,C_\Gamma^{-1/2}\,g_*(T_\mathrm{prod})^{1/4}.
\label{eq:relation_xi_decay_width}
\end{equation}
The last expression is only valid, if $g_*(T)\equiv g_*(T_\mathrm{prod})$ is constant during DM production.
As expected smaller decay widths give rise to a hotter second DM subset. The reason, why we use the half-life $t_{1/2}$ instead of $\Gamma^{-1}$, stems from the matching between a production assuming a shifted 
MB distribution and an explicit long-lived particle production mechanism, which will be explained in \cref{subsec:comparion_non_thm_prod}. Therefore, such long-lived decays are well approximated by a suitable choice of $\xi$ based on \cref{eq:relation_xi_decay_width} and we can use it as an input parameter for our simulations.
\section{Constraints on the model parameter space}
\label{sec:constraints_on_model_parameter}

Depending on the temperature ratio $\xi$, the subdominant production mechanism may lead to a warm or even hot DM subset which could lead to a significant contribution to the effective number of relativistic species, $N_\text{eff}$, or alter small scale structures. Therefore the 2TDM can be constrained by cosmological and astrophysical observations and measurements. Stringent constraints arise from flux spectra analyses of the Lyman-$\alpha$ forest and the number of dwarf galaxies of the Milky Way as well as the measured value for $N_\text{eff}$. 

\subsection{Limits from $N_\text{eff}$}
\begin{figure}[ht!]
 \centering
 \includegraphics[width=0.9\textwidth]{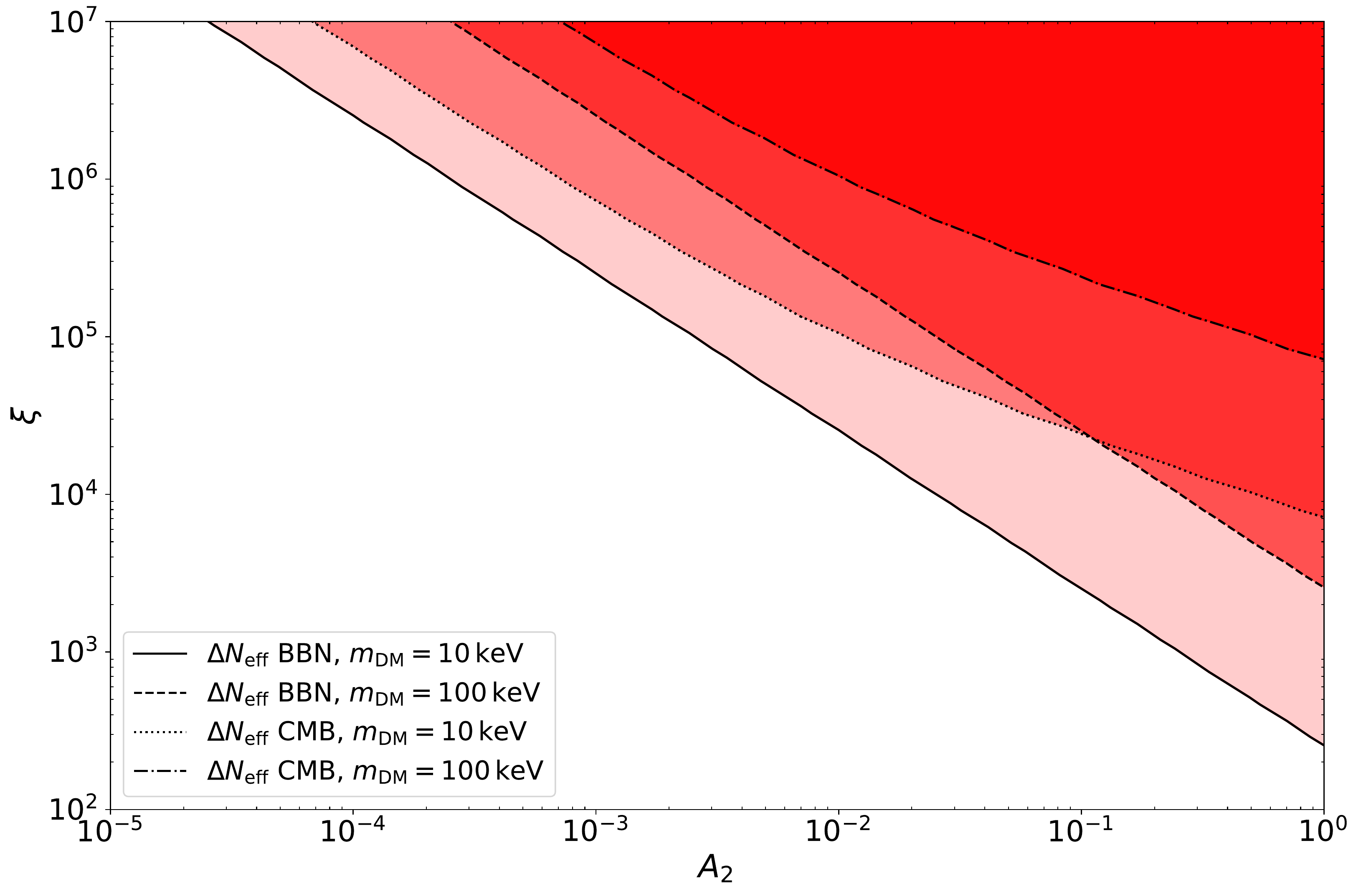}
 \caption{$\Delta N_\text{eff}$ bounds shown as red shaded regions and derived from \cref{eq:Neff_large_T} assuming $m_P=1\TeV$. The black solid line corresponds to $m_\mathrm{DM}=10\keV$ and the dashed line to $m_\mathrm{DM}=100\keV$ using the BBN bound $\Delta N_\text{eff}<0.35$. The dotted and dashed-dotted line are the corresponding bounds for $m_\mathrm{DM}=10\keV$ and $100\keV$ from CMB measurements, $\Delta N_\mathrm{eff}<0.28$. For $A_2=1$ temperature ratios $\xi\gtrsim 240$ are excluded.}
 \label{fig:N_eff_bounds}
\end{figure}

Hot DM effectively acts as radiation in the early universe and hence increases $N_\text{eff}$ by an amount $\Delta N_\text{eff}$. For the SM, $N_\text{eff}=3.044$ \cite{Akita:2020szl} while measurements by the Planck collaboration yield  $\displaystyle N_\text{eff} = 2.99^{+0.34}_{-0.33}$ ($95\%$ C.L.) (TT, TE, EE+lowE+lensing+BAO)  \cite{Planck:2018vyg} from the cosmic microwave background (CMB) whereas at the onset of big bang nucleosynthesis (BBN), $\displaystyle N_\text{eff} = 2.88\pm 0.52$ ($95\%$ C.L.) \cite{Pitrou:2018cgg}. Since BBN takes place at much earlier times the latter bound is more relevant for us, because the warm DM subset has more time to cool down until the CMB epoch.\\
We follow the procedure outlined in \cite{Merle_2015} and estimate $\Delta N_\text{eff}$ by comparing the kinetic energy of the DM species with temperature $T_2 = T_1 \xi$ to the energy density of a massless Dirac fermion with a temperature equal to the neutrino temperature $T_\nu$, which is given by \\$2\rho_\text{ferm}= \frac{7\pi^2}{60} T_\nu^4$:
\begin{align}
\label{eq:Neff_full}
\Delta N_\mathrm{eff} \equiv & \,  \frac{\rho(T_1)-n(T_1)\,m_{\text{DM}}}{2\rho_\text{ferm}} 
=  \, \frac{60}{7\pi^4}\left( \frac{T_1}{T_\nu} \right)^4\frac{m_{\text{DM}}}{T_1} \frac{g_*( T_1)}{g_*(m_P)} \times \notag \\
& \times \int\limits_0^\infty \dd z_1\, z_1^2 \left( \sqrt{1+ \left( \frac{g_*( T_1)}{g_*(m_P)} \right)^{2/3} \left( \frac{z_1 T_1}{m_{\text{DM}}} \right)^2}-1 \right)f(z_1 / \xi ), \\
\notag \mathrm{where} \quad z_1 = & \,x_1\left( \frac{g_*(m_P)}{g_*(m_P/r)}\right)^{1/3},
\end{align}
is a redefinition of the comoving momentum $x_1$ including $g_*(T)$. The dependence on the parameters $A_2$ and $\xi$ are encoded in $f(z_1/\xi)$.
The prefactor $(T_1/T_\nu)^4$ evaluates to $(11/4)^{4/3}$ below temperatures of $1\MeV$ and can be dropped for temperatures above. 
In the temperature range we are interested in, $T_1\approx 1\MeV$, one can simplify \cref{eq:Neff_full} by neglecting small expressions. Using the expression for $f(z_1/\xi)$ given in \cref{eq:equ_momentum_distr} we find in 
this case
\begin{align}
 \Delta N_\mathrm{eff} & \simeq \left( \frac{11}{4} \right)^{4/3}
 \frac{450}{7\pi^3}\, C_{\Gamma,2}  \left(\frac{g_*(T_1)}{g_*(m_P)} \right)^{4/3} \xi^4 \,.
\end{align}
This expression has to be evaluated at temperatures $T_1$ for given choices of $\xi$ and $A_2$ which are defined in $C_{\Gamma,2}$.
Inserting \cref{eq:Omega_h_combined} one can find an expression for $\Delta N_\mathrm{eff}$ only in terms of our model parameters
\begin{align}
	\Delta N_\mathrm{eff} \simeq
	3.1\cdot 10^{-3} 
	\left( \frac{g_*(T_1)}{10.75} \right)
	\left( \frac{ \Omega_\mathrm{DM}h^2}{0.12} \right)
	\left( \frac{10\keV}{m_\mathrm{DM}} \right)
	\left( \frac{g_*(T_1)/10.75}{g_*(m_P)/106.75} \right)^{1/3}
	\left( \frac{2}{g} \right) A_2\, \xi\,.
	\label{eq:Neff_large_T}
\end{align}
In \cref{fig:N_eff_bounds}, $\Delta N_\text{eff}$ bounds from BBN and CMB are compared for two different DM masses: the solid and dashed line represents BBN limits, i.e. we can use \cref{eq:Neff_large_T} evaluated at $T_1=1\MeV$, for $m_\text{DM}=10\keV$ and  $m_\text{DM}=100\keV$ respectively, 
while the dotted and dashed-dotted line correspond to CMB limits for the same masses. The later two are derived by evaluating \cref{eq:Neff_full} at $T_1 = 0.24\,\mathrm{eV}$. As already pointed out, the limits from BBN are in general stronger than the respective $\Delta N_\mathrm{eff}$ results from CMB, 
especially for smaller $\xi$ values. As such, the largest possible temperature $T_2$ for a $10\keV$ DM particle is $\sim 240\,T_1$, assuming $g=1$, i.e. a scalar DM species. For fermions or vector particles this bound has 
to be rescaled accordingly. Moreover, one can observe that the limits from CMB scale differently compared to BBN bounds at smaller temperatures. In that regime, expanding the square root in \cref{eq:Neff_full} 
leads to additional powers of $T_1$ and hence a larger temperature sensitivity.   

\subsection{Limits from structure formation}
\label{sec:limits_from_structure_formation}
\begin{figure}[ht!]
	\centering
	\includegraphics[width=0.8\textwidth]{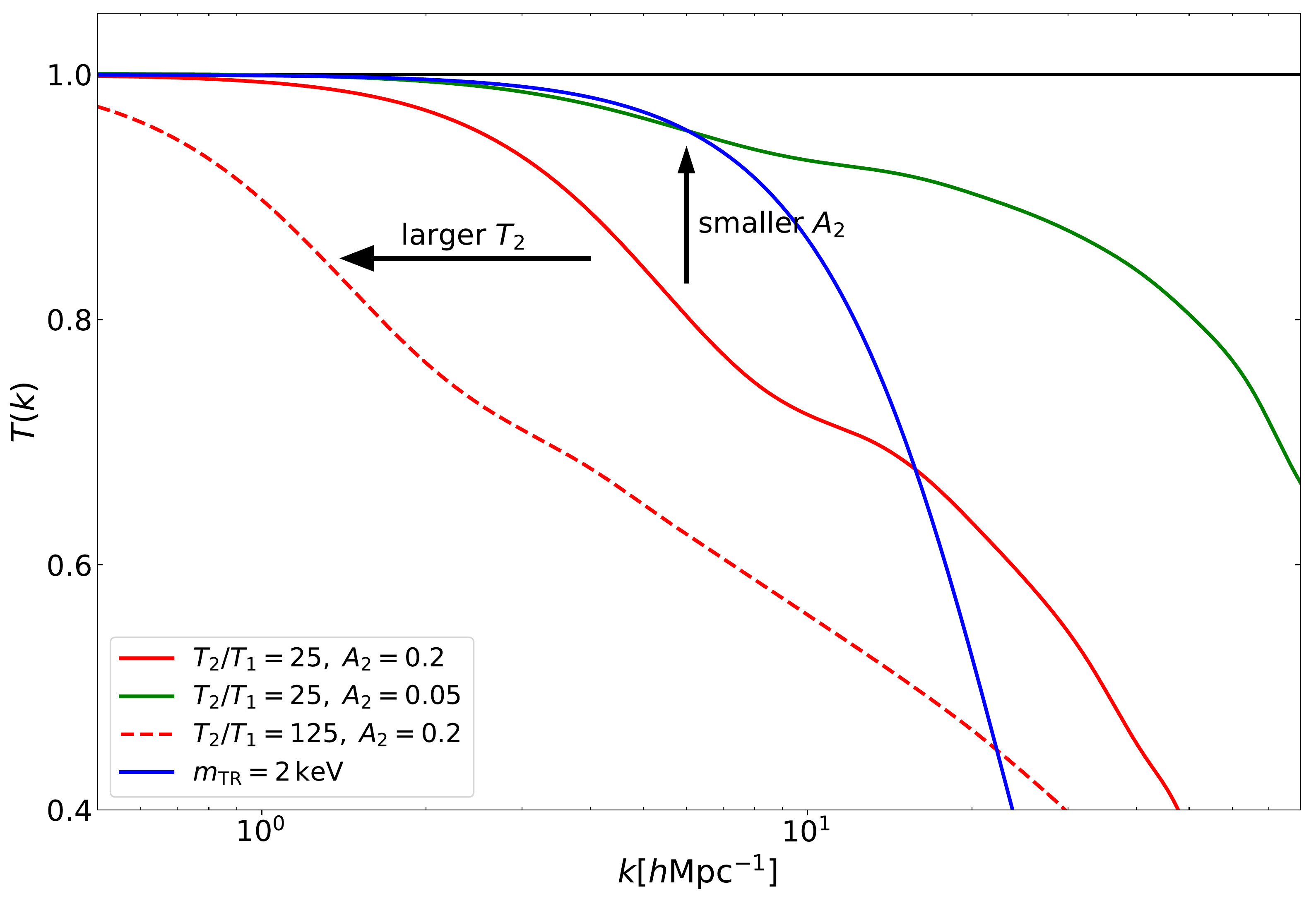}
	\caption{Transfer function $T(k)$ for a 2TDM model where $m_\text{DM}=30\keV$, $A_2 = 0.2$ and $\xi = 25,\;125$ shown in red and dashed red respectively, while the blue line is derived from \cref{eq:transfer_func_ana} for a thermal relic mass $m_\text{TR}=2\keV$ indicating a potential limit from an analysis on structure formation. The green line correspond to $\xi=25$ and a smaller $A_2=0.05$. Applying a half-mode analysis, all three parameter choices would be excluded, but limits from observables on the matter power spectrum only exclude the two red lines, while the green line is not in conflict.}
	\label{fig:transfer_func_example}
\end{figure}

Generally, a detailed study for a given warm DM or mixed warm/hot and CDM model (in the following we will refer to these as WDM) would require hydrodynamical $N$-body simulations to infer their impact on the formation of cosmological structures. 
However, the influence of a specific model on small scales can usually be understood by comparing its corresponding matter power spectrum with the associated power spectrum of $\Lambda$CDM. Based on this comparison, conclusions can be drawn whether a given WDM model features a too large suppression of structure formation at small scales.
We use the public code \texttt{CLASS} \cite{Lesgourgues:2011rh,Lesgourgues:2011re} to derive the matter power spectrum for the 2TDM model. For this, we calculate the DM momentum distribution function given in \cref{eq:DM_momentum_distr_func} for different $x$ values and forward the result as a data table to \texttt{CLASS}. Then our DM model can be specified in the program by using its ``non-cold DM" component.\\
Generally the suppression features of WDM models can be parameterized in terms of the transfer function $T(k)$
\begin{equation}
 T(k)^2 =  \frac{P_{\text{WDM}}}{P_{\Lambda \text{CDM}}}.
\end{equation}
Assuming for a moment that DM is composed of only one thermal relic with mass $m_\text{TR}$, $T(k)$ can be expressed via the following analytic expression \cite{Viel:2005qj}
\begin{align}
T(k) & =\left(1+(\alpha k)^{2\beta}\right)^{-5/\beta},
\label{eq:transfer_func_ana}\\
\text{where}\quad \beta & = 1.12,\;
\alpha = 0.049\left(\frac{m_\text{TR}}{1\keV}\right)^{-1.11} \left(\frac{\Omega_\text{TR}}{0.25}\right)^{0.11}\left(\frac{h}{0.7}\right)^{1.22}h^{-1} \text{Mpc}\,.\notag
\end{align}
Given that many limits stemming from structure formation are quoted in terms of $m_\text{TR}$ one can employ a half-mode analysis (see for instance \cite{Baumholzer:2020hvx}) to match these limits with specific WDM models.
On the contrary, this procedure is not suitable for the 2TDM model. Similar to mixed hot and cold DM models we are dealing with a plateau in the transfer function \cite{Hu:1997mj,Boyarsky:2008xj} and as such a simple half-mode analysis does not capture the whole picture of this model. Some examples are shown in \cref{fig:transfer_func_example} where transfer functions for three different parameter choices with $m_\text{DM}=30\keV$ are compared to a thermal relic with mass $m_\text{TR}=2\keV$ shown in blue. The solid and dashed red curves correspond to $A_2 = 0.2$ and $\xi = 25$ or $125$ respectively and the green curve has $A_2 = 0.05$ and $\xi=25$.  While the transfer functions shown in red and green are generally smaller than the thermal reference below some scale $k$ due to the warmer DM subset, they still cross the blue line because the larger first DM subset features a milder suppression of scales. Using a half-mode analysis would therefore exclude all three parameter choices and even a pretty small deviation for $A_2\ll 1$ would be disfavored by such analysis. However, the parameter choice shown in green is still allowed by limits on structure formation observables.
As such we are going to use the matter power spectrum directly to extract limits on the model parameter space.

\subsubsection{Lyman-$\alpha$ forest}
\label{subsubsec:lyman_alpha_forest}
The so called ``\Lya" is a way to examine the intergalactic medium (IGM) by looking at absorption lines from neutral hydrogen along the line-of-sight of highly redshifted quasars. It is a good tool to constrain WDM models, since it measures deviations from $\Lambda$CDM at smallest scales based on the distribution of the IGM.\\
Lyman-$\alpha$ forest surveys probe the spectra of quasars at redshifts $3 <z < 5$ and the corresponding flux power spectrum $P_F(z,k_\nu)$ for different scales in velocity-space, $k_\nu$, and redshifts is derived. In principle, one would have to set up a full $N$-body simulation for a specific WDM model and fit the flux power spectrum directly to the data to derive constraints, but there is a shortcut to this procedure, using the one-dimensional power spectrum $P^{1D}(k)$ (see for instance \cite{Schneider:2016uqi,Kobayashi:2017jcf,Murgia:2017lwo}).\\
The relation between the flux power spectrum and the one-dimensional spectrum is given by a bias function $b^2(k) \equiv P_F(k_\nu,z)/P^{1D}(k)$. The conversion factor between velocity-space scales and inverse comoving length scales used in the power spectrum is given by $\displaystyle k=\frac{H(z)}{1+z}k_\nu$. The one-dimensional power spectrum is the momentum integral of the matter power spectrum evaluated at $z=0$ and given by
\begin{equation}
P^{1D}(k) = \frac{1}{2\pi}\int\limits_{k}^{\infty}\dd k^\prime\, k^\prime P(k^\prime)\,.
\end{equation}
In practice, an upper cutoff for the scale $k$ has to be used and in the following $k=200\,\mathrm{h/Mpc}$ will be employed.
In a similar fashion as before, the deviation for a given WDM model is parameterized by defining the ratio
\begin{equation}
\phi(k) = \frac{P^{1D}_\mathrm{WDM}(k)}{P^{1D}_{\Lambda\text{CDM}}(k)},
\label{eq:xi_ratio}
\end{equation}
where $P^{1D}_{\Lambda\text{CDM}}(k)$ is the one-dimensional power spectrum of $\Lambda$CDM. 
The next step is to quantify how much a given WDM model differs from a $\Lambda$CDM scenario.
So one integrates \cref{eq:xi_ratio} over all scales typically probed by Lyman-$\alpha$ observations given in the range $(k_\text{min},k_\text{max})$
\begin{equation}
A= \int\limits_{k_\text{min}}^{k_\text{max}}\dd k\, \phi(k)\,,
\end{equation}
and this quantity can be used to approximate the amount of suppression for the 2TDM model by defining the estimator
\begin{equation}
\delta A \equiv\, \frac{A_{\Lambda \text{CDM}}-A}{A_{\Lambda \text{CDM}}}\,,
\label{eq:delta_A}
\end{equation}
where $A_{\Lambda \text{CDM}}=k_\text{max}-k_\text{min}$. The results we are using are from an analysis examining the combination of the MIKE/HIRES and the XQ-100 datasets \cite{Irsic:2017ixq}. MIKE/HIRES observed quasars with redshifts $z=4.2$--$5.4$, while XQ-100 measured between $z=3$--$4.2$. Both sets combined span a range in $k_\nu$-space from $(0.003$--$0.08)$\,s\,$\text{km}^{-1}$. Hence, we will set $k_\text{min}=0.5\,\text{h/Mpc}$ and $k_\text{max}=10\,\text{h/Mpc}$ in the following. 
To derive limits on the 2TDM, we have to define a reference WDM model with a corresponding $\delta A_\text{ref}$ value first. The analysis in \cite{Irsic:2017ixq} yields a lower bound for thermal WDM given by $m_\text{WDM}=3.5\keV$ (at 95\% C.L.) considering a conservative thermal history of the universe. Under the assumption of a power-law evolution this bound strengthens to $m_\text{WDM}=5.3\keV$. Using these masses as input parameters for a thermal WDM model we derive the following values
\begin{align}
m_\text{WDM}=3.5\keV & \quad  \Rightarrow\quad \delta A_\text{ref,1} = 0.30,\\
m_\text{WDM}=5.3\keV & \quad  \Rightarrow\quad \delta A_\text{ref,2} = 0.20.
\end{align}
That means all parameter points in our scenario which have $\delta A > \delta A_\text{ref}$ are excluded since their scale suppression is too strong.

\subsubsection{Number of Milky Way satellites}
\label{subsubsec:number_of_MW_satellites}
\begin{figure}[ht!]
	\centering
	\includegraphics[width=0.8\textwidth]{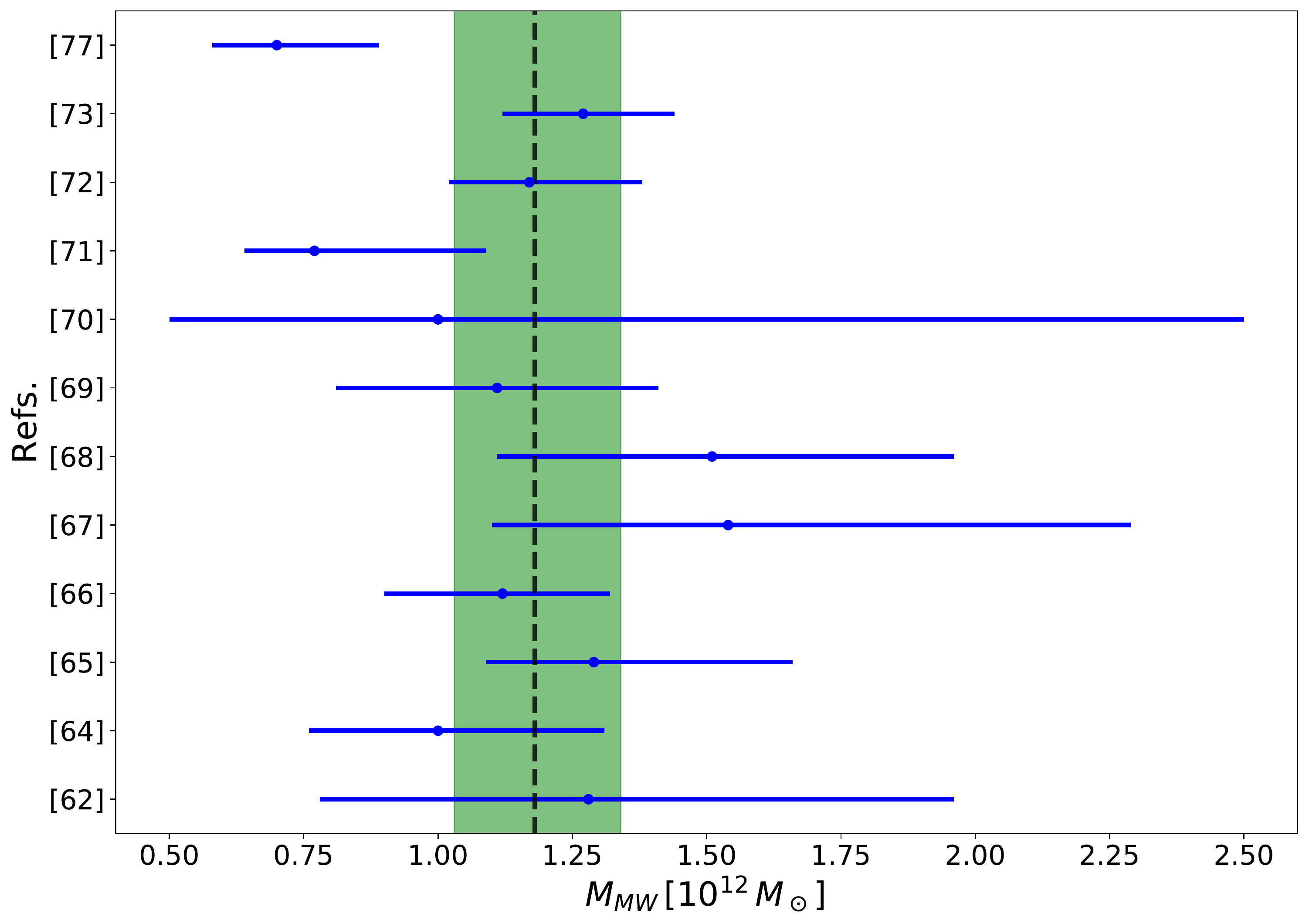}
	\caption{Compilation of different MW mass analyses using recent GAIA DR2 data. Further information on the results can be found in \cite{Monari_2018,Deason_2019,Grand_2019,2020MNRAS.494.4291C,Watkins_2019,Fritz_2020,Posti_2019,Vasiliev_2019,Eadie_2019,Callingham_2019,Li_2020,Necib:2021vxr}. The black dotted line is the combination of all measurements and the green shaded region gives a $2\sigma$ error range. On the contrary, blue error bars correspond to 68\% C.L. limits of the respective analyses.}
	\label{fig:MW_plot}
\end{figure}

Generally, WDM models predict less satellites for MW like galaxies compared to vanilla $\Lambda$CDM, because they tend to suppress the mass distribution function of the subhalos \cite{Colombi:1995ze, Viel:2005qj, Boyarsky:2008xj, Benson_2012, Schneider:2014rda}. 
Similarly, the 2TDM model might lead to a too large suppression of the subhalos and so one can use the observed number to constrain the model parameters by calculating the corresponding number of subhalos.
\\ 
Before we start to address the issue how to count the number of MW companions, we will present an analytic result for the predicted number of subhalos $N_\text{sub}$ for specific WDM models. In \cite{Schneider:2013ria,Schneider:2014rda,Schneider:2016uqi} the authors derived a formula to estimate $N_\text{sub}$ for a given matter power spectrum 
\begin{equation}
\frac{\dd N_\text{sub}}{\dd M_\text{sub}}= \, \frac{1}{C} \, \frac{1}{6\pi^2} \, \frac{M_\text{MW}}{M_\text{sub}^2} \, \frac{P(1/R_\text{sub})}{R_\text{sub}^3\sqrt{2\pi(S_\text{sub}-S_\text{MW})}}\,.
\label{eq:number_sub_halos}
\end{equation}
Here, $M_\text{sub}$ and $M_\text{MW}$ denotes the mass of the subhalo and the MW, respectively. $C$ is a normalization constant used to match with $N$-body simulations and depends on the definition of the host halo. In our case, the boundary of the host halo is set by the criterion that its density is 200 times the critical density $\rho_c$ of the universe and hence we use $C=34$ in the following. The variance $S_i$ of the amount of subhalos and the scaling between mass $M_i$ and radii $R_i$ of the subhalos or the MW galaxy are given by
\begin{equation}
S_i=\frac{1}{2\pi^2}\int\limits_0^{1/R_i}\dd k\,k^2P(k)\,, \quad M_i =\frac{4\pi}{3} \, \Omega_m \, \rho_c \, (2.5R_i)^3\,,
\end{equation}
where the matter density is given by $\Omega_m = 0.315$ \cite{Planck:2018vyg}.
Integrating \cref{eq:number_sub_halos} from $10^8 M_\odot/h$ to $M_\text{MW}$ yields the number of subhalos for a given parameter point in our scenario with an associated power spectrum $P(k)$. 
\\
There are two uncertain numbers in the following discussion: first, the observed amount of MW subhalos and second the mass of the MW. Addressing the counting of subhalos we follow the approach outlined in \cite{Schneider:2016uqi,Murgia:2017lwo}: there exist 11 ``classical" satellites. They are combined with 15 ultra-faint satellites found by SDSS. Those number is multiplied by a factor of 3.5 because of the limited sky coverage of SDSS. In total this yields $N_\text{sub}=64$. One should note that in addition to SDSS several more ultra-faint satellites or satellite candidates have been reported (see for instance \cite{DES:2019vzn}) by several other surveys, and as such we think of $N_\text{sub}$ as a conservative estimate of MW companions.
In fact, simulations \cite{Newton:2020cog,10.1093/mnras/sty1085} predict $\mathcal{O}(100)$ subhalos which could be detected with future observations  and therefore open new possibilities to further test our scenario.\\
An estimation of the MW mass is done in \cite{Wang:2015ala,McMillan_2016,Callingham:2018vcf,2020MNRAS.494.4291C,Karukes:2019jwa,Wang:2019ubx,Dekker:2021scf}, and it is found to be in in range between $1\times 10^{12}\, M_\odot/h <M_\text{MW}<2\times 10^{12}\,M_\odot/h$. 
With the second data release of the GAIA mission, several works have calculated the MW mass using different analysis techniques (see \cite{Monari_2018,Deason_2019,Grand_2019,2020MNRAS.494.4291C,Watkins_2019,Fritz_2020,Posti_2019,Vasiliev_2019,Eadie_2019,Callingham_2019,Li_2020} and \cite{Wang:2019ubx} for an overview) and a compilation of these results is shown in \cref{fig:MW_plot}. Combining every measurements following the procedure outlined in \cite{Barlow:2004wg} we find for the MW mass
\begin{equation}
	 M_\text{MW}=\,1.18^{0.16}_{-0.15}\times 10^{12} M_\odot\; \mathrm{(95\% C.L.).}
	 \label{eq:MW_mass_choice}
\end{equation}
In the following, we will take the limits as a lower (i.e. \textit{light}) and upper (i.e. \textit{heavy}) MW mass bound and reject parameter points if they have $N_\text{sub}<64$.\\
\\
As a side remark, we comment briefly on the MW mass dependence of this procedure. We matched it to the prediction of the \textit{Aquarius} simulation (\cite{Lovell:2013ola}, taking $M_\mathrm{sub}>10^8 \,M_\odot$) which is $N_\text{sub}=158$ by calculating $P(k)$ for $\Lambda$CDM and a larger MW mass, $M_\text{MW}\simeq 2\times 10^{12}\, M_\odot$. In contrast, using the mass choices of \cref{eq:MW_mass_choice}, the number of subhalos yields only $\mathcal{O}(100)$ in the $\Lambda$CDM case.\footnote{This observation was already pointed out in \cite{2012} as a possible explanation for the former missing satellites problem.}

\section{Detailed study of the parametrization}
\label{sec:going_beyond_thermal}
So far, we were assuming that the parent particles are thermalized when decaying, but in general this assumption does not hold for rather long-lived or weakly coupled particles. 
In the following, we show that a shifted MB distribution for $f(x,r)$ can be used to describe the momentum distribution of DM produced from the decay of non-thermal parent particles, whose distribution function is set by a freeze-in or freeze-out mechanism.
Further, we study the impact of a temperature dependent $g_*(T)$ on the DM momentum distribution.

\subsection{Non-thermalized parent particles}
\label{subsec:comparion_non_thm_prod}
\begin{figure}[ht!]
	\centering
	\includegraphics[width=0.6\textwidth]{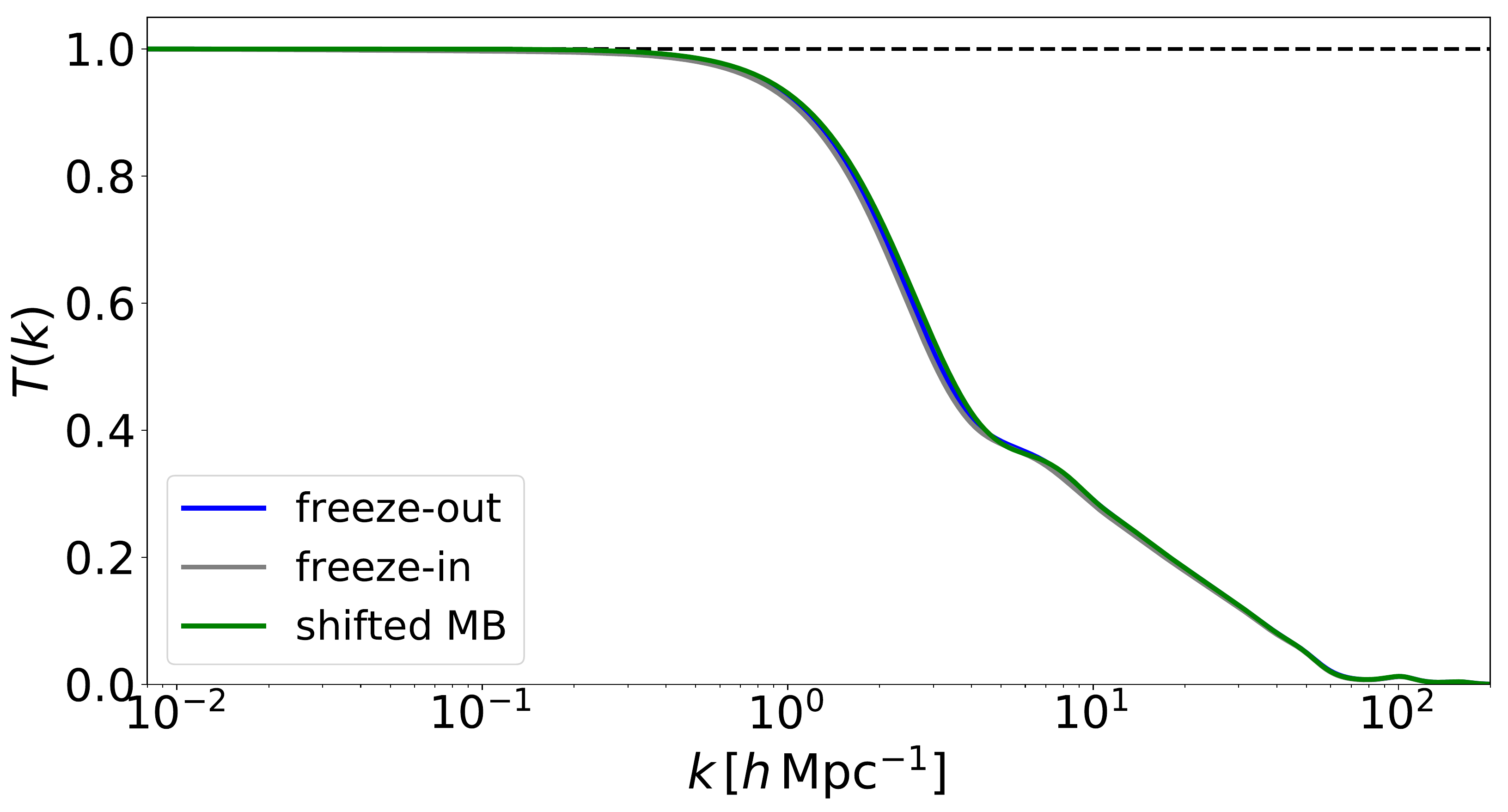}
	\caption{
		Comparison of the transfer function $T(k)$ for a shifted MB distribution and a freeze-in/freeze-out scenario where the temperature ratio is set to $\xi =  40$ and $A_2 = 0.5$, while $m_\mathrm{DM}=10\keV$. We see that the analytical shifted MB distribution (green curve) is a good tool to approximate the numerical results for decays of frozen-out or frozen-in parents, shown in blue and gray, respectively. Hence we will use it in the following analysis to deduce constraints from structure formation. The horizontal black dashed line corresponds to a pure $\Lambda$CDM scenario.   
	}
	\label{fig:fN_comparison_prod_mechanisms}
\end{figure}

In \cref{subsec:analytical_modeling} we suggested to use a shifted MB distribution to model late time decays of parent particles. In the following, we are going to verify that this is a good approximation for the cases where $f_P(x,r)$ is determined by a freeze-out, if the particle is sufficiently coupled or, if not, by a freeze-in.
\\
To compare this approximation and the two late time regimes we make use of 
\cref{eq:relation_xi_decay_width} to mock these decays, which are governed by $C_\Gamma$, with a shifted MB distribution and corresponding temperature ratio $\xi$ (see \cref{eq:shifted_MB_distr}). This will guarantee that DM is produced at approximately the 
same time. For the case of the frozen-in or frozen-out parent particle, we calculate $f(x,r)$ numerically by inserting the corresponding $f_P(x,r)$ into \cref{eq:master_equation}. \\
As an illustrative example we choose $C_\Gamma \simeq 5.2 \cdot 10^{-4}$, which corresponds to $\xi \approx 40$ and set $A_2=0.5$ for each production mechanism, to compare the results for the corresponding transfer functions
for the case of a shifted MB distribution with the other two cases in \cref{fig:fN_comparison_prod_mechanisms}. Shown in green is the result using a shifted MB distribution, while the transfer function for DM production by decays of parent particles after they are frozen-out or frozen-in are shown in blue and gray, respectively. It can be observed that the corresponding matter power spectra feature a similar scale where they deviate from $\Lambda$CDM. Only decays of frozen-in parent particles give rise to a slightly earlier drop in $T(k)$. Overall, the deviation between an appropriately shifted MB distribution and a freeze-in or freeze-out parent is only marginal; this allows us to model late time decays using our simpler analytic expressions.
\\
To highlight implications for structure formation even more, we calculate $N_\text{sub}$ using the lighter MW mass and $\delta A$ for fixed $m_\text{DM}=50\keV$ and different choices of $A_2$. The respective results are shown in \cref{tab:SF_comparison}.
\begin{table}[ht!]
	\centering
	\tabulinesep=1.1mm
	\begin{tabu}{|c|c|c|c|c|c|c|}
		\hline
		& \multicolumn{2}{c|}{shifted MB} & \multicolumn{2}{c|}{freeze-in} & \multicolumn{2}{c|}{freeze-out}\\
		\hline
		$A_2$	&	$N_\text{sub}$	&	$\delta_A$	&	$N_\text{sub}$	&	$\delta_A$	&	$N_\text{sub}$	&	$\delta_A$\\
		\Xhline{1pt}
		0.1	&	76	&	0.335	&	76	&	0.341	&	76	&	0.334	\\	
		0.3	&	37	&	0.627	&	36	&	0.638	&	37	&	0.630	\\
		0.5	&	14	&	0.787	&	14	&	0.798	&	14	&	0.791 \\
		\hline
	\end{tabu}
	\caption{Comparison of the corresponding structure formation observables using a shifted MB distribution for parent particles and the respective momentum distributions for non-thermal parent particles. The DM mass is set to $50\keV$ and only the abundance $A_2$ is varied. The predictions for $N_\text{sub}$ and $\delta_A$ are nearly identical.}
	\label{tab:SF_comparison}
\end{table}
\\
In summary, our findings indicate that we can model late time decays to a good approximation by a temperature shifted momentum distribution assuming thermalized parents only.
\\
As a final remark, one can use above mentioned methods to place absolute lower mass bounds on the DM mass, $m_\mathrm{DM}^\mathrm{lim}$, by assuming $\xi =1$. These values act as a guideline for the allowed parameter choices for the 2TDM and the respective limits from structure formation are summarized in \cref{tab:mass_limit}.
\begin{table}[ht!]
	\centering
	\tabulinesep=1.1mm
	\begin{tabu}{|c|c|c|c|c|}
		\hline
		&
		 \multicolumn{2}{c|}{$N_\mathrm{sub}$} & \multicolumn{2}{c|}{Lyman-$\alpha$} \\
		\hline
		 	&	light MW	&	heavy MW	&	$\delta A_{\mathrm{ref},1}$	&	$\delta A_{\mathrm{ref},2}$\\
		 	\hline
		$m_\mathrm{DM}^\mathrm{lim}\,[\keV]$ &	12.8	&	9.0	&	12.7	&	7.7\\
		\hline
	\end{tabu}
	\caption{Lower DM mass limit $m_\mathrm{DM}^\mathrm{lim}$ using constraints from structure formation assuming $\xi = 1$, i.e. all of DM has a common temperature $T_2 = T_1$.}
	\label{tab:mass_limit}
\end{table}

\subsection{Impact of a variation in $g_*(T)$ during DM production}
\label{sec:modifications_variable_dof}
\begin{figure}
  \centering
  \includegraphics[width=0.75\textwidth]{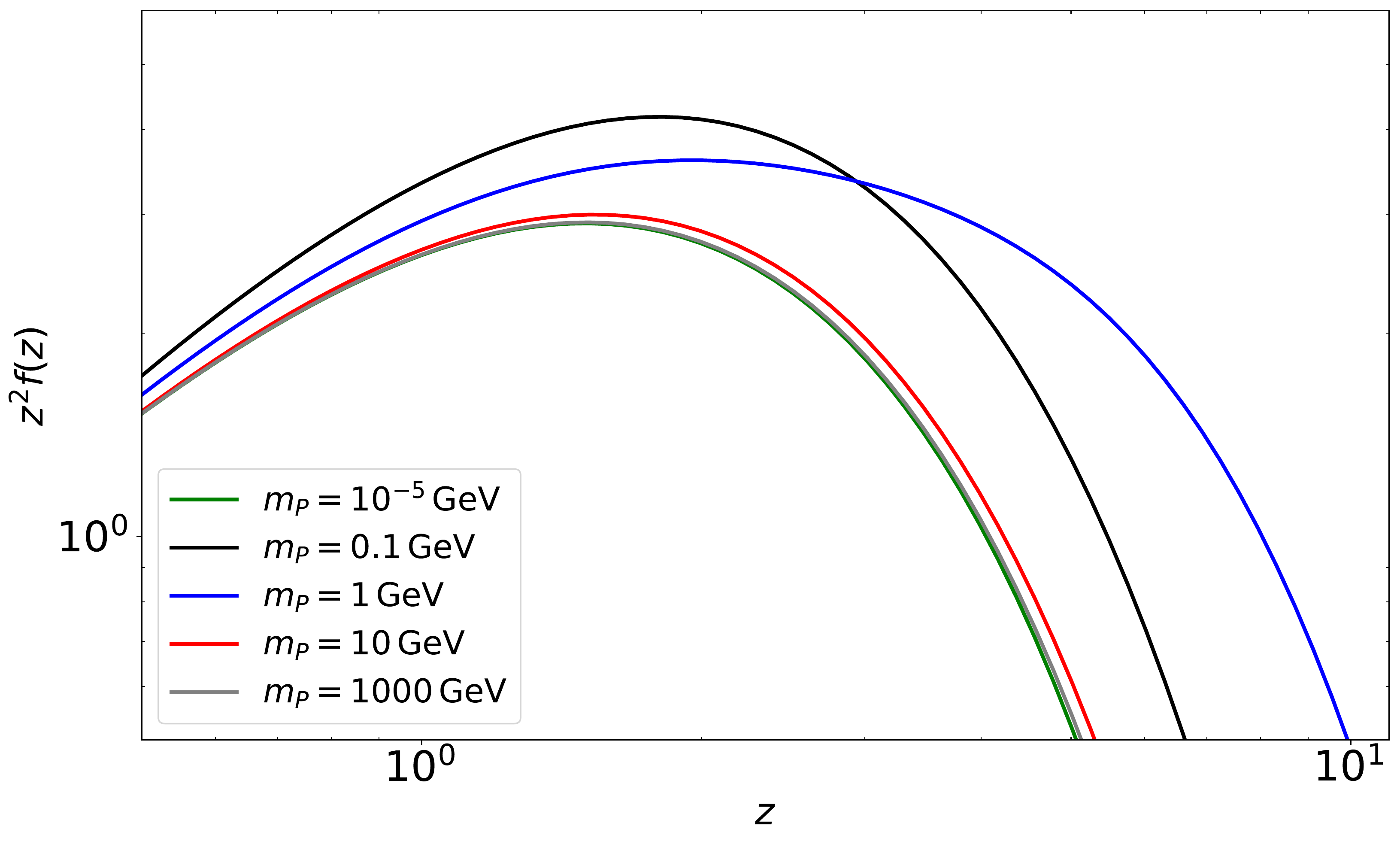}
  \captionof{figure}{Numerical results for $z^2f(z)$ as defined in \cref{eq:boltzman_varying_dof} for different masses of the parent particle, ranging from $m_P=10^{-5}\GeV$ to $m_P=1\TeV$ and using \cref{eq:equ_momentum_distr} for $f(z)$ as an illustration.
Masses of $0.1$ and $1\GeV$ lead to the biggest impact on the shape of the spectrum, due to the rapid change of $g_*(T)$ around those times, while the distributions for other parent particle masses have similar shapes. All distributions are rescaled with a factor $\sqrt{g_*(m_P)}$ for an easier comparison.}
 \label{fig:Changing_dof_momentum_distr_rescaled_2}
\end{figure}

In the previous sections we have treated the number of entropic degrees of freedom $g_*(T)$ as a fixed quantity. This assumption is only well justified for high decoupling temperatures,
$T_\text{dec}\gtrsim 160\GeV$, where $g_*(T)=106.75$ is constant (neglecting non-SM degrees of freedom). As such, this simplification may be applicable for the first DM subset, but this simplifying assumption does not hold necessarily for the second warmer DM subset. Of course, the impact of a varying $g_*(T)$ depends on the production time of the second DM subset, which is related to the mass of its respective parent particle.\footnote{Freeze-in
is most dominant at temperatures $r\approx 3$, i.e. $T \approx m_P/3$.} In the following, we are using analytical expressions for $g_*(T)$ given in the appendix of \cite{Wantz:2009it}. Introducing a new variable $z$ for the comoving momentum
\begin{equation}
z = \left( \frac{g_*(m_P)}{g_*(m_P/r)} \right)^{1/3} x\,,
\end{equation}
one can rewrite \cref{eq:master_equation} to derive the more general DM momentum distribution function \cite{Konig:2016dzg}
\begin{align}
\frac{\partial f(z, r)}{\partial r} = &  \,2\, \frac{C_\Gamma}{\sqrt{g_*(m_P/r)}}  \left(1 - \frac{r\,\partial_r g_*(m_P/r)}{3g_*(m_P/r)}\right) \frac{r^2}{z^2} \left( \frac{g_*(m_P)}{g_*(m_P/r)} \right)^{2/3} \notag \\
& \times \int\limits_{y_\mathrm{min}}^\infty \dd y \, \frac{r^2}{\sqrt{r^2+y^2}} \,f_P(y,r)\,,
\label{eq:boltzman_varying_dof}\\
\notag \mathrm{where}\quad y_\mathrm{min} = & \, z\left(\frac{g_*(m_P/r)}{g_*(m_P)}\right)^{1/3} - \frac{r^2}{4z}\left(\frac{g_*(m_P)}{g_*(m_P/r)}\right)^{1/3}.
\end{align}
To outline the impact of a variation in $g_*(T)$, we insert the momentum distribution from \cref{eq:equ_momentum_distr} into \cref{eq:boltzman_varying_dof} and vary $m_P$. Assuming $\xi=1$, the results for $z^2f(z)$ are shown in \cref{fig:Changing_dof_momentum_distr_rescaled_2}. The mass scale $m_P$ sets the time of the DM production, i.e. DM is produced earliest for $m_P = 1\TeV$. For a better comparison we rescale the momentum distribution with a factor $\sqrt{g_*(m_P)}$, to compensate for the decrease in $g_*(T)$ at late times. 
Besides from this overall change in magnitude, the shape of the distributions are going to change when DM is produced during periods of time where $g_*(T)$ is rapidly changing: while the curves derived for $m_P = 10^{-5}\GeV$, $10\GeV$ and $1\TeV$ are nearly identical, the other two curves, where $m_P = 10^{-1}\GeV$ and $1\GeV$ clearly deviate, because the  QCD phase transition leads to a rapid change in $g_*(T)$ at $T\approx 200\MeV$. In particular, the momentum distribution
gets shifted to larger $z$ values when DM is produced during this period of time, as can be seen from the blue curve, for which $m_P = 1\GeV$. 
\\
To summarize, one has to be careful when defining a proper temperature ratio $\xi$, because its definition is done by using $\langle  z \rangle = 2.5$ assuming constant $g_*(T)$.\footnote{In the case $g_*(T) = \mathrm{const}$, $z\equiv x.$} That means one has to include a shift in the averaged momentum due to a change in $g_*(T)$ when comparing against our results shown in the next section. To quantify the required shift we compare the result for $\langle z \rangle$ of \cref{eq:boltzman_varying_dof} to the reference case \cref{eq:master_equation} for $m_P$ between $1\MeV$ and roughly $1\TeV$ and $\xi$ up to $\simeq 6000$. A maximum deviation of $\approx 2.5$ can be observed at large $\xi$ values and for $m_P\simeq 1\MeV$, see \cref{fig:pt_ratio} in the appendix. 
For a given parent particle mass, one can extract a function $h(\xi)$ from this contour plot and rescale the temperature ratio accordingly, $\xi^\prime = \xi / h(\xi)$. 
\\
A change in the degrees of freedom also leads to a heating of the photon plasma compared to the decoupled DM temperature. This does not affect the ratio $T_2/T_1$ but for the derivation of the matter power spectrum the DM temperature has to be defined with respect to the photon temperature $T_\gamma$. 
Compared to this reference temperature, the DM temperatures evolve as: 
\begin{equation}
	\frac{T_1}{T_\gamma} = \left( \frac{g_*(T_\gamma)}{g_*(T_{\mathrm{prod},1})} \right)^{1/3}, \quad \frac{T_2}{T_\gamma} = \xi\left( \frac{g_*(T_\gamma)}{g_*(T_{\mathrm{prod},2})} \right)^{1/3}.
	\label{eq:temperature_relative_photon_bath}
\end{equation} 
Late times of production will come with a decrease in the number of entropic degrees of freedom, $g_*(T_{\mathrm{prod},2}) < g_*(T_{\mathrm{prod},1})$ and compared to the photon temperature, 
$T_2$ is increased and $\xi$ is larger by a factor $\left(g_*(T_{\mathrm{prod},1})/ g_*(T_{\mathrm{prod},2})\right)^{1/3}$. 
This can be taken into account by defining a shifted temperature ratio $\xi^\prime = \xi  \left(g_*(T_{\mathrm{prod},1})/ g_*(T_{\mathrm{prod},2})\right)^{-1/3}$ to include the reheating effect of the thermal plasma.\\
We will explain these rescaling procedures in more detail in \cref{sec:application_toy_models} where we apply it to specific models and extract limits on the allowed temperature ratio.

\subsection{Three-body decays}
Compared to two-body decays, decays involving three or more particles are more likely to feature small decay widths, because they can be suppressed by powers of small couplings, heavy off-shell intermediate particles or large mass ratios. In case of three-body decays one can have production of a DM particle $X$ via the processes $P\to S\,S\,X$, $P\to S\,X\,X$ or $P\to X\,X\,X$.
Similarly to the previously discussed two-body decays one can derive an analytic expression for the DM momentum distribution, assuming a thermalized parent particle and $m_P \gg m_S, m_X$~\cite{DEramo:2020gpr}:
\begin{equation}
f(x) \propto  x^{-1.2} \exp\left( -1.11x \right).
\label{eq:3body_momentum_distr}
\end{equation}
The prefactor of this function is fixed by demanding $A_2 = \int \dd x_1 (x_1)^2 f(x_2)$. 
In contrast to two-body decays, the energy of the parent particle is distributed among three particles. This has two consequences for the interpretation of our results in the next section. 
First, the averaged momentum should be smaller by a factor of $2/3$, in fact, we found that $\langle x \rangle=1.62$ using \cref{eq:3body_momentum_distr}. Further, the same factor has to be used when mapping 
the assumed decay width to the temperature ratio, in that case that relation is given by $\displaystyle\xi \simeq 0.16/\sqrt{C_\Gamma}g_*(T_\mathrm{prod})^{1/4}$ for the case of constant $g_*(T)$ during DM production.
We have checked that both, using the momentum distribution given in \cref{eq:equ_momentum_distr} with a specific choice for $\xi$ and \cref{eq:3body_momentum_distr} with an appropriately rescaled ratio, give rise to nearly identical matter power spectra.
\\
Although we do not present analytical results for three-body decays of frozen-out or frozen-in parent particles, we are confident that one can make use of our procedure to extract limits for 
the case when the second subset is produced via late three-body decays with appropriately chosen values for the temperature ratio $\xi$. Since the deviations in the matter power spectrum are not very drastic, we expect 
that the results from \cref{fig:pt_ratio} holds for three-body decays to a good degree and this allows to derive limits on production via three-body decays by applying our findings. We leave the derivation of a full picture 
of more general three-body decays for future work.

\section{Analytical fitting of the exclusion limits}
\label{sec:fit_results}
\begin{figure}[ht!]
	\centering
	\includegraphics[width=0.9\textwidth]{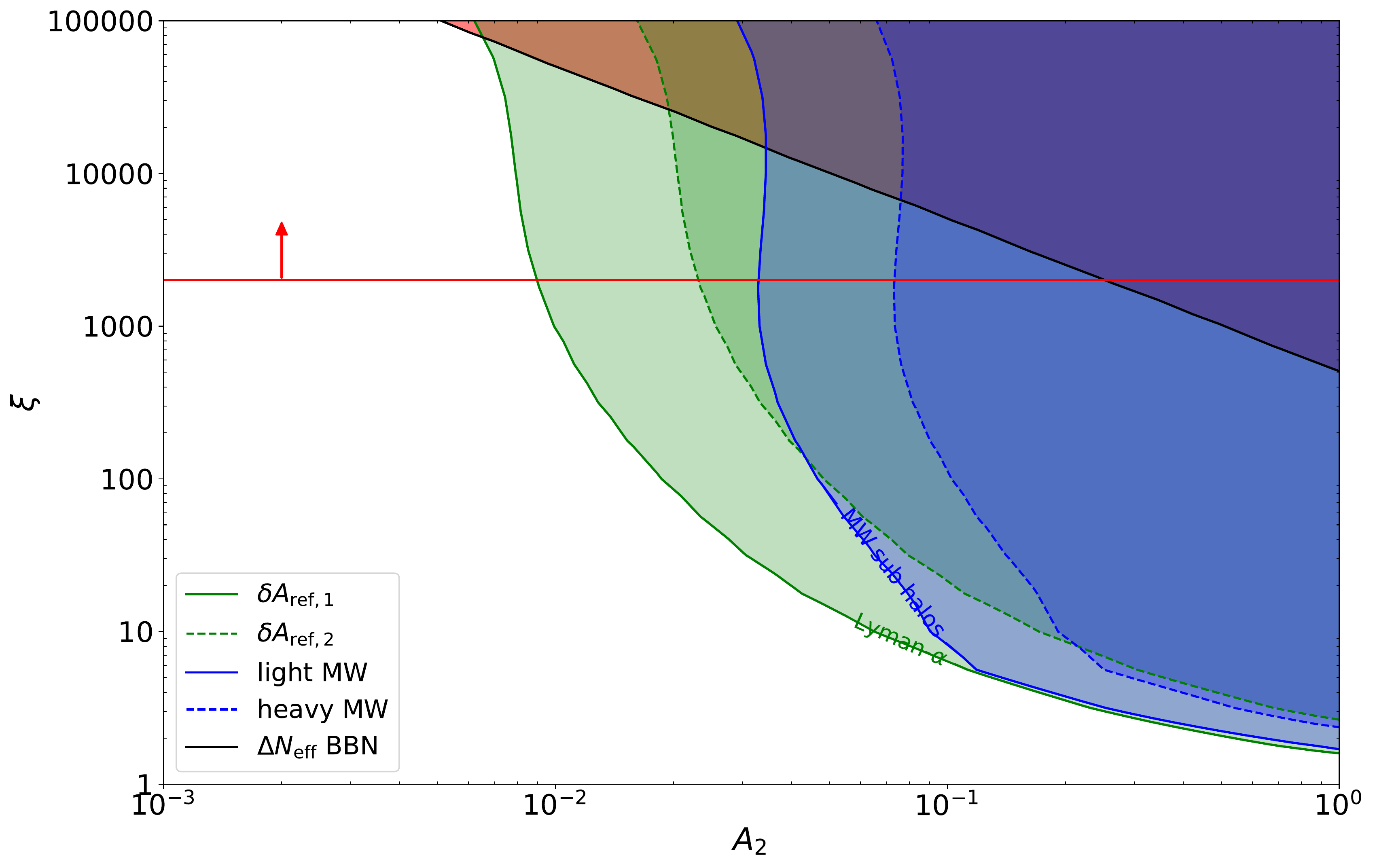}
	\caption{Limits from structure formation and the number of entropic degrees of freedom for $m_\text{DM}=20\keV$ and fixed $g_*(T)=106.75$. The blue shaded region is disfavored by the MW subhalo count and the green shaded region by limits from Lyman-$\alpha$ surveys, respectively. The solid lines are the corresponding stronger limits, whereas the weaker constraints are shown as dashed lines. See text for more details on these bounds. 
	The red shaded region in the upper right corner is disfavored by a too large $\Delta N_\text{eff}$ value. Finally, the red line indicates the region of parameter space, where our structure formation simulation setup becomes less reliable.}
	\label{fig:SF_comb_20keV}
\end{figure}

In the following we are using the tools discussed in \cref{sec:constraints_on_model_parameter} to answer the question how large and how hot the second DM subset can be. We are going to present our results in terms of the $A_2$--$\xi$ 
parameter space of the 2TDM. Results are derived for different choices of $m_\text{DM}$ as higher DM masses give rise to weaker constraints. 
\\
As an example we show the constraints on the parameter plane in \cref{fig:SF_comb_20keV} where we set $m_\text{DM}=20\keV$ and keep $g_*(T) = 106.75$ fixed until all of the DM production has been completed.
As can be seen the limits from structure formation place strong constraints on the temperature ratio $\xi$ in the range $10^{-2} < A_2 < 1.0$ while the $\Delta N_\mathrm{eff}$ bound from the BBN epoch starts to become relevant at rather large temperature ratios, $\xi > 10^4$. Above this value, the bounds from structure formation become less reliable, because the hot DM subset starts to act like dark radiation instead of matter, an effect not captured in the calculation of $P(k)$. 
\\
In the following, our aim is to provide our results in a model-independent way such that they can be applied to a variety of scenarios. 
For this reason we fit the respective exclusion limits with an exponential of the form $\xi(A_2) = \exp( p_0\cdot A_2^{-p_1} + p_2)$, which we found to be generally suitable. Furthermore, one parameter can be removed, because we know that the curve endpoint, $\xi(A_2=1)$, scales linearly with the DM mass starting from $m_\mathrm{DM}^\mathrm{lim}$ as given in \cref{tab:mass_limit}.
By eliminating $p_2$, the exponential can be reduced to the following expression
\begin{equation}
\xi(A_2) = \frac{m_\mathrm{DM}}{m_\mathrm{DM}^\mathrm{lim}}\exp\left[ p_0 \left( A_2^{-p_1} -1 \right)\right]\,.
\label{eq:fit_equation}
\end{equation}
This enables us to use either the abundance $A_2$ or the temperature $\xi$ as an input parameter and derive constraints on the other variable. The fitting parameter results for all four exclusion contours for $m_\mathrm{DM} = 20\keV$ are shown in \cref{tab:fit_para_20keV}.
\begin{table}[ht!]
	\centering
	\tabulinesep=1.1mm
	\begin{tabu}{|c|c|c|c|c|}
		\hline
		&	\multicolumn{2}{c|}{Lyman-$\alpha$}	&	\multicolumn{2}{c|}{$N_\mathrm{sub}$ } \\
		\hline	
		&	$\delta A_{\mathrm{ref},1}$	&	$\delta A_{\mathrm{ref},2}$	&	\textit{light} MW	&	\textit{heavy} MW	\\
		\hline
		$p_0$	&	$0.464$	&	$0.546$	&	$0.141$	&	$0.196$	\\
		$p_1$	&	$0.581$	&	$0.672$	&	$1.11$	&	$1.29$	\\
		\hline
	\end{tabu}
\caption{Fit results for respective exclusion limits based on \cref{eq:fit_equation} using $m_\mathrm{DM}=20\keV$.}
\label{tab:fit_para_20keV}
\end{table}
\\
While the endpoint of the limit scales linearly with DM mass, the exclusion curve for $A_2 < 1$ changes non-trivially. Therefore we extend this fitting procedure for other DM masses and simulate the exclusion limits for $m_\mathrm{DM}$ between $20\keV$ and $500\keV$. Then, the respective parameters $p_i$ are extracted and fitted using the following polynomial
\begin{equation}
p_i\left( \frac{m_\mathrm{DM}}{\mathrm{keV}} \right) = a_i + b_i\left( \frac{m_\mathrm{DM}}{\mathrm{keV}} \right)^{-1} + c_i\left( \frac{m_\mathrm{DM}}{\mathrm{keV}} \right) + d_i \left( \frac{m_\mathrm{DM}}{\mathrm{keV}} \right)^2,
\label{eq:poly_fit}
\end{equation}
to derive a final fit function which takes $A_2$ and $m_\mathrm{DM}$ as input parameters to give 
the allowed temperature ratio $\xi\equiv \xi(m_\mathrm{DM},A_2)$. The results for all 8 fit parameters can be found in \cref{app:details_fit_pi} in the appendix.
We explicitly compare this analytical fit against numerical simulations in the context of toy model examples in \cref{subsec:thermal_and_out_of_equi_DM_const_g}.\\
\\
The assumption that $g_*(T)$ is constant during DM production obviously does not hold in general. We will explain in the following (see also \cref{fig:flow_chart} in appendix) how to adapt our results beyond the simplified picture and including a change in $g_*(T)$ as well.
\\
The starting point is to collect the $A_2$--$\xi$ relation given in \cref{eq:fit_equation}, the expression for $p_i(m_\mathrm{DM})$ (see \cref{eq:poly_fit}) and the corresponding fit parameters given in \cref{tab:fit_results_pi}. As mentioned, this gives a first approximation of the exclusion limits, under the assumption $g_*(T) = \mathrm{const}=106.75$ during DM production. If this does not apply for the second subset, because its production happens at times $T_{\mathrm{prod},2}$ where $g_*(T)$ is changing, it will lead in general to a warmer DM subset as compared to the case where $g_*(T)$ is constant. Two corrections have to be done: first, the exclusion limit on $\xi$ has to be divided by $\left( g_*(T_{\mathrm{prod},1}) / g_*(T_{\mathrm{prod},2}) \right)^{1/3}$. Second, one has to take the change in $\langle x \rangle$ into account by extracting a correction function $h(\xi)$ for the corresponding $m_P$ from \cref{fig:pt_ratio}.
This gives a rescaled version of the temperature ratio, $\xi^\prime$
\begin{equation}
\xi^\prime = \frac{\xi(m_\mathrm{DM},A_2)}{h\big(\xi(m_\mathrm{DM},A_2)\big)}\left( \frac{g_*(T_{\mathrm{prod},2})}{106.75} \right)^{1/3}.
\label{eq:xi_shifted}
\end{equation}
Additionally, if the first subset features $\langle x \rangle$ different from $2.5$ or has $g_*(T_{\mathrm{prod},1}) < 106.75$, which gives rise to a higher DM temperature $T_1$ relative to the photon bath, one further step has to be done before the corresponding limits on $\xi$--$A_2$ can be extracted. 
This change of $A_1$ can be quantified by multiplying $\langle x \rangle$ with a factor $\alpha$, which is either given by the ratio between the averaged momentum and our reference case, $\alpha = \langle x \rangle / 2.5$, or by the entropy dilution factor, $\alpha = (106.75/g_*(T_{\mathrm{prod},1}))^{1/3}$.
Regarding its matter power spectrum, a DM with temperature $\alpha \, T_1$ and mass $m_\mathrm{DM}$ has the same properties as a DM with temperature $T_1$ and mass $m_\mathrm{DM} / \alpha$. However, changing the mass by $1/\alpha$ in \cref{eq:poly_fit}, one has to rescale the outcome for $\xi^\prime$ by multiplying it with $\alpha$, because the second DM subset is not affected.
Taking all these corrections into account, the temperature limit is set by
\begin{equation}
\xi^\prime = \alpha \, \frac{\xi(m_\mathrm{DM}/\alpha,A_2)}{h\big(\xi(m_\mathrm{DM},A_2)\big)}\left( \frac{g_*(T_{\mathrm{prod},2})}{106.75} \right)^{1/3}.
\label{eq:xi_shifted_alpha}
\end{equation}
\\ 
In the next section we will explain in more detail how to incorporate these effects using several toy models as examples. A special emphasis will be put on the proper extraction of the limits on $\xi$ when $g_*(T)$ is not fixed.

\section{Application to toy models}
\label{sec:application_toy_models}
To explain the matching between our parametrization and ``real" DM models we consider several toy models where the DM is produced in different ways. We derive constraints on the model parameter space using the analytical fits to the structure formation constraints that we introduced in the previous section, and compare them to the limits one obtains by importing the dark matter momentum distribution in CLASS and running the full simulation. As can be seen below, in most cases our analytic fit provides a very good approximation to the actual limit. 
\\
The first two toy models are combinations of thermal freeze-in~\cite{Hall:2009bx} and Super-WIMP mechanisms~\cite{Covi:1999ty,Feng:2003uy}, which appears naturally as soon as the dark sector has more than the minimal particle content, see e.g.~\cite{Hooper:2011aj, Heeck:2017xbu, Garny:2018ali, Feng:2003uy,Baumholzer:2018sfb,Baumholzer:2019twf,Decant:2021mhj,Parimbelli:2021mtp,Dienes:2020bmn, Dienes:2021itb}. While the masses for the first example are chosen such that the number of relativistic degrees of freedom doesn't change between the production times, the second case shows how to incorporate the appropriate corrections in more realistic scenarios where $g_*(T)$ does vary. The third toy model considers the special case that one of the parent particles itself is not thermalised.

\subsection{Toy model I: thermalized + out of equilibrium parents, constant $g_*(T)$}
\label{subsec:thermal_and_out_of_equi_DM_const_g}
\begin{figure}[ht!]
	\centering
	\includegraphics[width=0.495\textwidth]{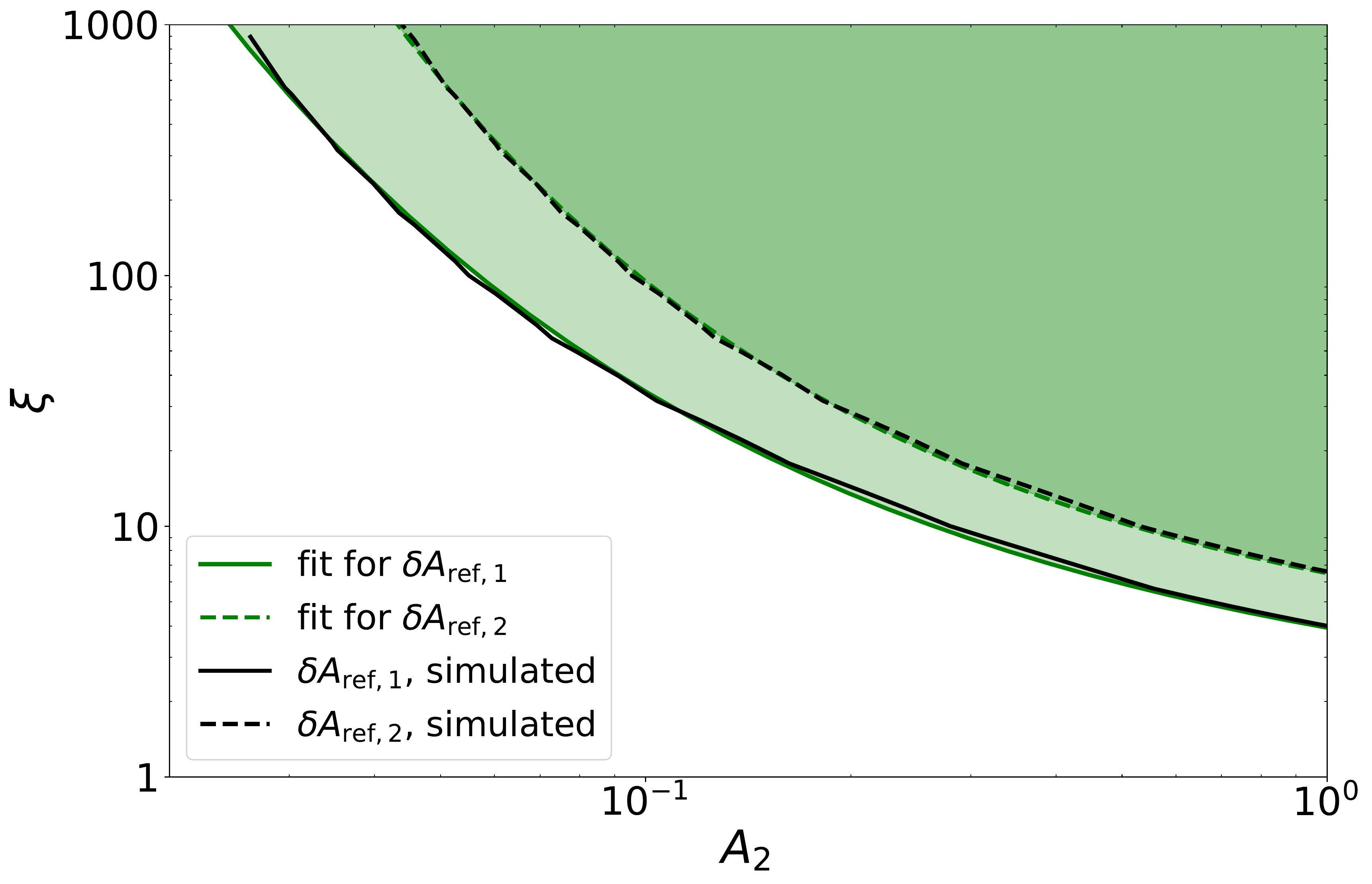}
	\includegraphics[width=0.495\textwidth]{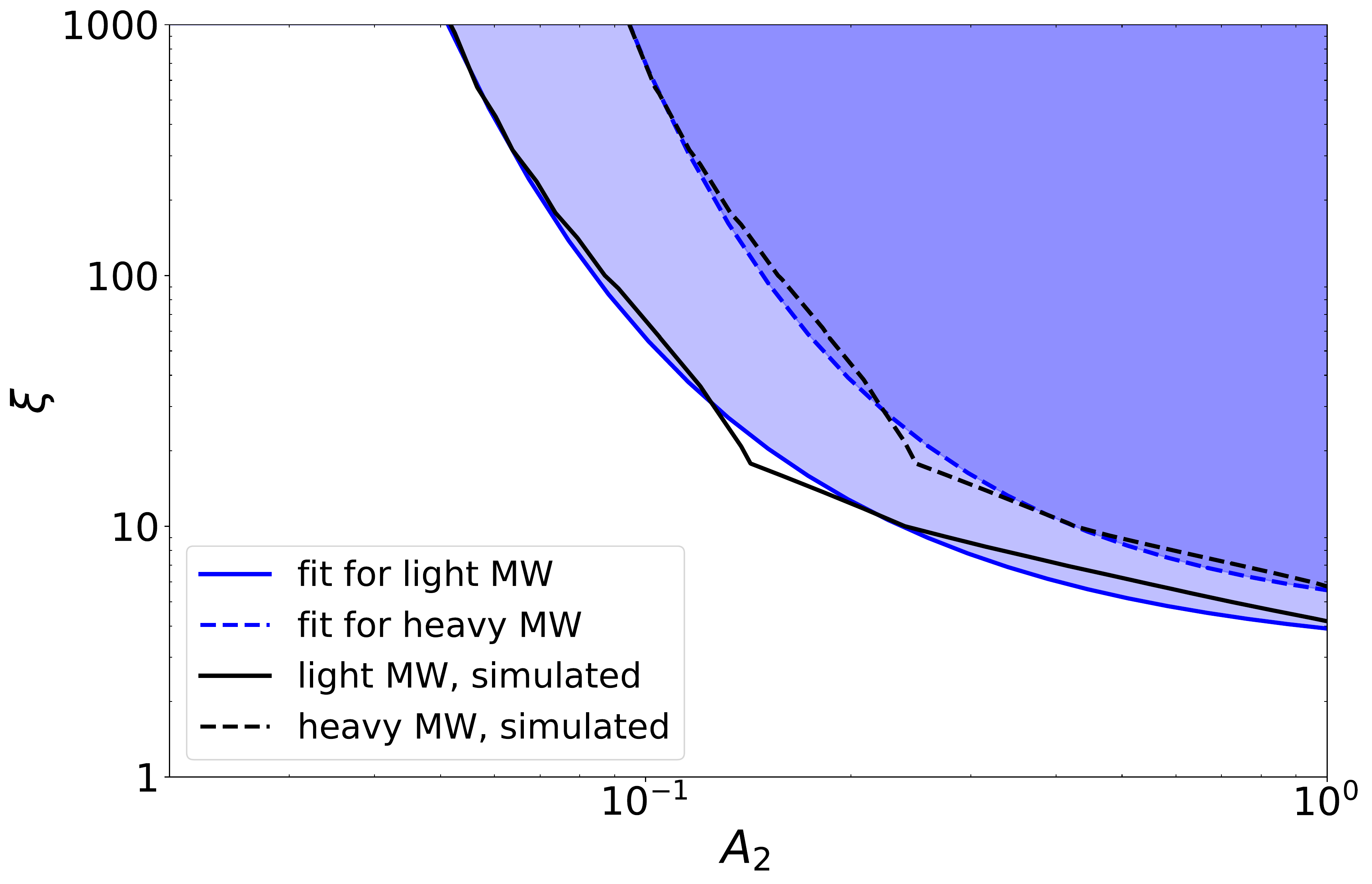}
	\caption{Structure formation limits given by shaded regions and derived from our analytic fit, \cref{eq:fit_equation,eq:poly_fit}, using $\delta A$ in the left and $N_\mathrm{sub}$ in the right figure. The DM mass is given by $m_\mathrm{DM}=50\keV$ and we assume
	$g_*(T)=\mathrm{const}$ during DM production. Shown as black lines are corresponding limits from a numerical simulation and solid/dashed lines correspond to stronger or weaker bounds, respectively.
	}
	\label{fig:limits_50keV_toy_model}
\end{figure}

We start by discussing two thermalized parent particles, $S$ and $P$, with masses $m_S = m_P = 50\TeV$ which are producing a DM species 
with mass $m_\mathrm{DM} = 50\keV$
with their respective decays. Due to their large mass the number of entropic degrees of freedom can be treated as constant until all of them decayed into DM. Only afterwards, the dilution of $g_*(T)$ has to be taken into 
account. We assume that $S$ decays rapidly, while remaining in equilibrium with the thermal plasma. This is going to produce an amount $A_1<1$ of DM particles with averaged momenta given by $\langle x \rangle = 2.5$. Additionally, $P$ is going to decay at late times, after it is already frozen-out and produces an amount
$A_2$ of DM, such that $A_1 + A_2=1$. 
Now, because these decays are taking place at later times compared to production via $S$ decays, one ends up with DM which is highly energetic compared to the thermal plasma. The difference in the respective DM temperatures gives the ratio $\xi$.
\\
The interesting question now is how hot the DM share $A_2$ can be, without being in conflict with observations of structure formation. Following the procedure outlined in the previous section, we can neglect modifications
stemming from a change in $g_*(T)$ during DM production and extract the limits on $\xi$ directly from \cref{eq:fit_equation}. 
\\
In \cref{fig:limits_50keV_toy_model} we compare our fitted exclusion limits for a DM mass of $50\keV$ against the limits obtained from a full numerical simulation. Here, the simulated results are derived passing the corresponding $f(x)$ into \texttt{CLASS}. The numerical results for $P(k)$ are then used to study the suppression effect on small scale structures as explained in \cref{sec:limits_from_structure_formation}. The left figure shows results from Lyman-$\alpha$ data (green shaded regions) and the right one from a MW subhalo count (blue shaded regions). In both figures, simulated results are shown as black lines and the stronger/weaker bounds are given by solid or dashed lines, respectively. As one can see, our fit gives a good approximation of the simulated results.  

\subsection{Toy model II: thermalized + out of equilibrium parents}
\label{subsec:thermal_and_out_of_equi_DM}
\begin{figure}[ht!]
	\centering
	\includegraphics[width=0.495\textwidth]{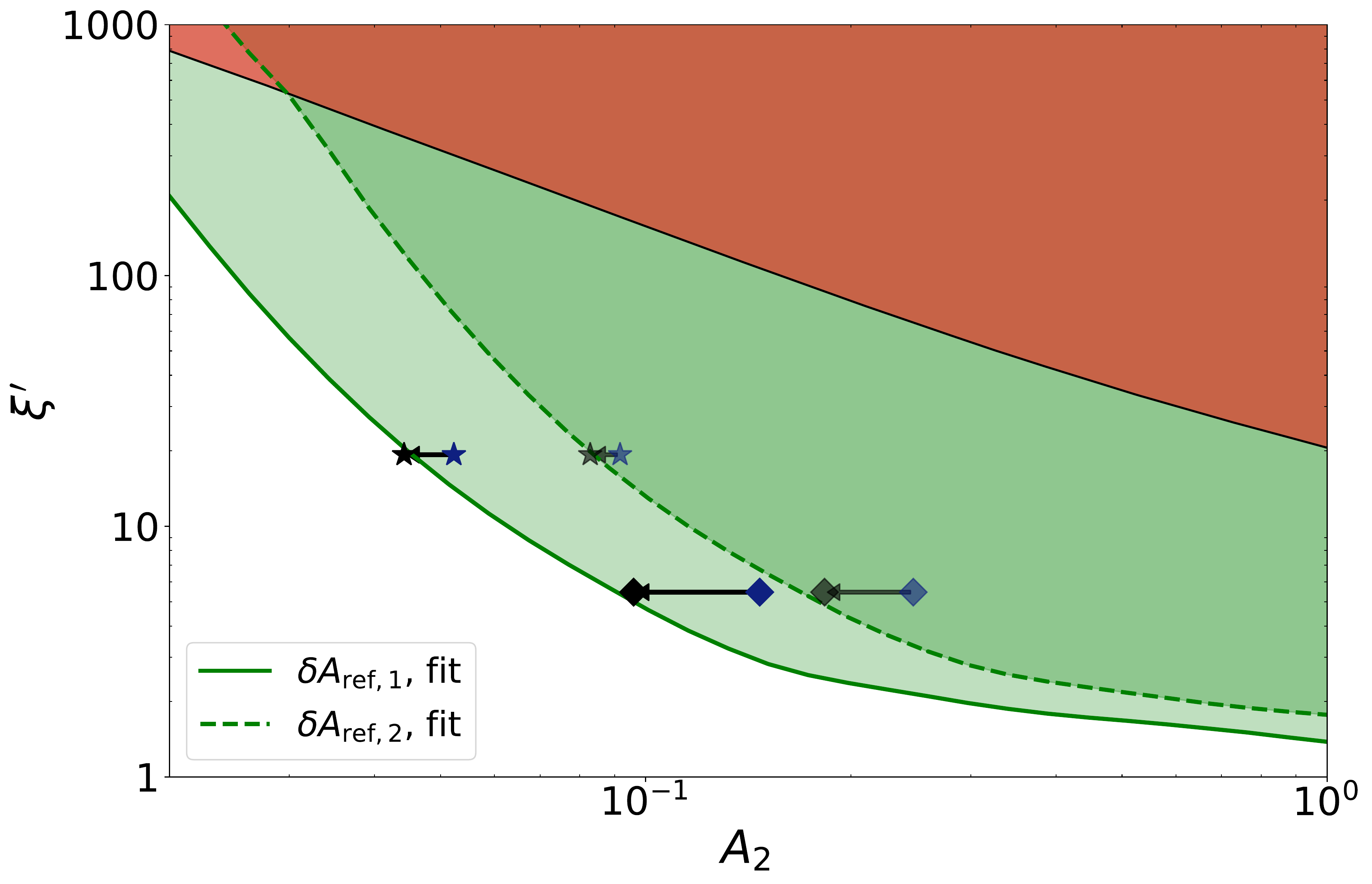}
	\includegraphics[width=0.495\textwidth]{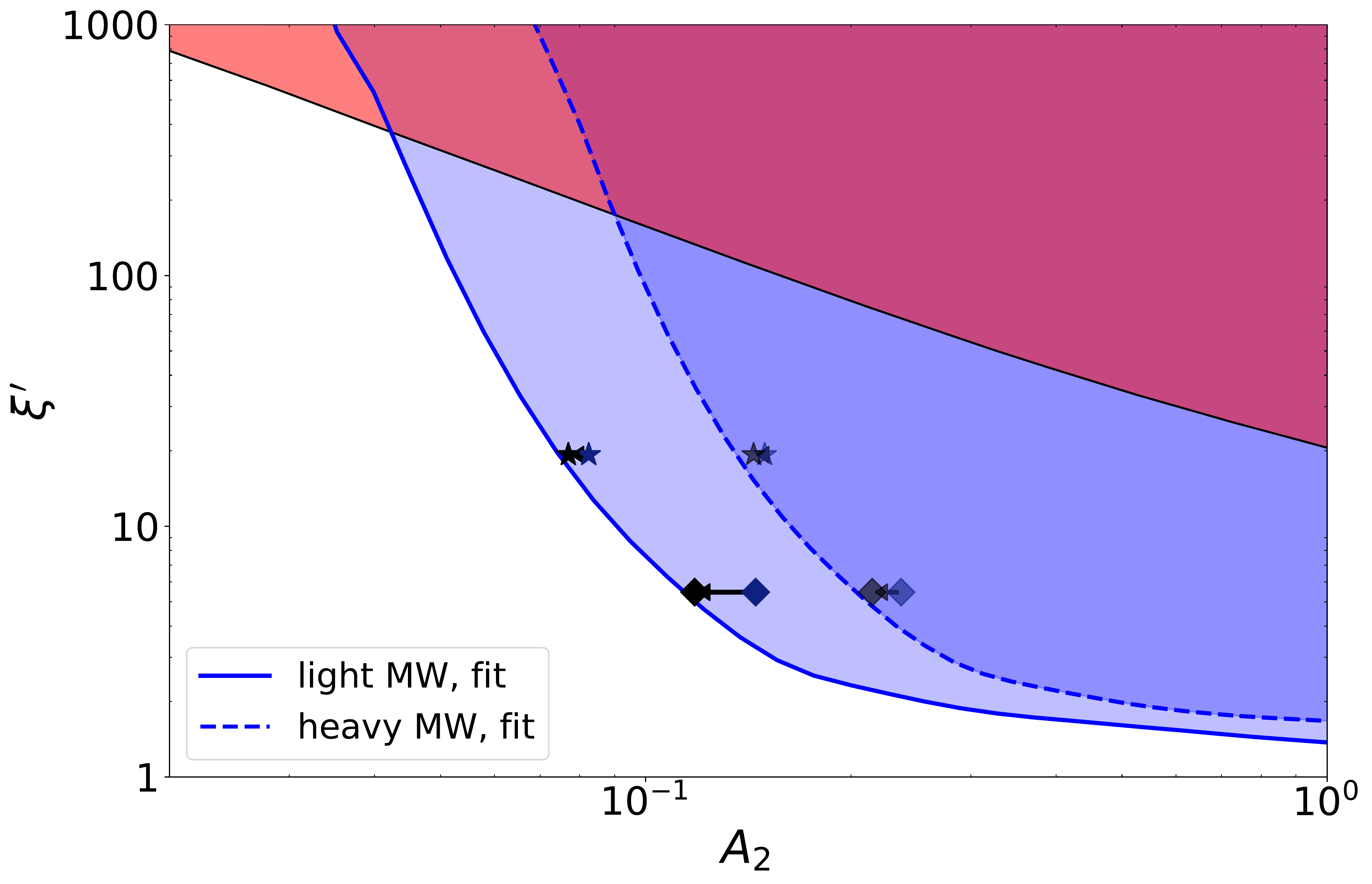}
	\caption{Structure formation limits using the $\delta A_{\mathrm{ref}}$ criterion in the left and a MW subhalo count in the right figure for the second toy model. Everything indicated by the green or blue shaded region is constrained by structure formation and solid lines correspond to strong and dashed lines to weak limits. For comparison, full numerical simulations to extract limits were run for $\xi = 10$ and $40$ and are indicated by the stars and diamonds in both plots. The difference between the purple and black symbols indicated by an arrow are due to specifics of this model choice and are further explained in the text. The red shaded region yields a too large $\Delta N_\mathrm{eff}$ stemming from the second subset.}
	\label{fig:limits_35keV_toy_model}
\end{figure}

In the previous example we illustrated how to interpret and extract the corresponding limits on $A_2$ and $\xi$ using our fit procedure under the assumption of constant $g_*(T)$ during the production of DM. 
Based on this, we consider now a similar setup as before: again, we have two thermalized parent particles, $S$ and $P$ whose decays will produce a DM species with mass $m_\mathrm{DM}=35\keV$. However, now the parent 
masses are given by $m_S = 1\TeV$ and $m_P=10\GeV$. As before, $S$ produces a DM amount $A_1$ while remaining in equilibrium with the thermal plasma, whereas an additional DM subset, $A_2$, is produced by late time decays of $P$ after it is frozen-out.\\
Since DM production now happens at times where $g_*(T)$ is changing, we have to compensate for this effect by using \cref{eq:xi_shifted}.
Here, we assume that the first subset is produced at early times where all SM particles are still part of the thermal bath, while the function $h(\xi)$ can be read off \cref{fig:pt_ratio} for the particle masses involved.
Now we can take $\xi^\prime$ as the constraint on the temperature ratio for this toy model. $\Delta N_\mathrm{eff}<0.35$ limits are indicated by the red shaded region and bounds derived using $\delta A_{\mathrm{ref},1}$ and light MW masses are shown in the left and right figure of \cref{fig:limits_35keV_toy_model} as green and blue solid lines respectively, while the respective weaker bounds are shown as dashed lines. Compared to \cref{fig:limits_50keV_toy_model} one can observe that the exclusion bands feature a kink around $A_2 = 0.2$ and therefore smaller $\xi^\prime$ values are excluded in this region. The reason is that at this point, $P$ particles decay around a temperature of $T\simeq 1\GeV$ where $g_*(T)$ is rapidly changing and hence it gives rise ro a larger averaged momentum $\langle z \rangle$, as illustrated in \cref{fig:Changing_dof_momentum_distr_rescaled_2}.
\\
To demonstrate the usefulness of our approach, we simulate the combined matter power spectrum for this toy model for two benchmark points with $\xi = 10$ and $40$. We explicitly insert the out of equilibrium momentum distribution for $S$ (see \cref{eq:momentum_distr_out_of_equi}) taking changes in 
$g_*(T)$ into account as well. The limits on $A_2$ for these choices are indicated by a star and a diamond in both plots: here, the slightly grayed out symbols indicate the weaker limits. These benchmark points have to be matched onto $\xi^\prime$ as well, by multiplying them with a factor $(106.75/g_*(T_{\mathrm{prod},2}))^{1/3}$, which evaluates roughly to $0.48$ for $\xi = 40$ and $0.55$ for $\xi = 10$.
\\
Further, it is noticeable that the benchmark points, indicated by purple stars or diamonds, seem to yield weaker constraints compared to our fit. For the scenario in mind, this can be explained by a DM fraction which is already produced while the parent particle is still in thermal equilibrium, giving rise to a peak at smaller momenta similar to the case of thermalized parent particle decays.
Instead our parametrization assumes that the second subset is fully produced via late time decays. Consequently, at large $\xi$ this fraction is only marginal, but becomes more dominant for smaller $\xi$. We calculated this fraction explicitly and numerically extracted the updated bound on $A_2$ shown as the black diamonds and stars in the plot. The difference between both is indicated by a black arrow and one can see, that the latter, more careful treatment fits better with our analytical result. It is interesting to note, that for this toy model, limits from $N_\mathrm{eff}$ competes with the respective weaker structure formation limits at small $A_2$ and large $\xi^\prime$.  
\\
Furthermore, we show how the constraints on $A_2$--$\xi$  can be matched onto specific parameters for a concrete freeze-out model. The abundance $A_2$, produced by late time decays, is fixed by the 
abundance of $P$ which in turn is set by the time of its freeze-out, $r_\mathrm{FO}$. Integrating \cref{eq:momentum_distr_out_of_equi} over $x$, the yield $Y_\mathrm{DM} = n_\mathrm{DM}/s$ is given by
\begin{align}
   A_2 = \frac{\Omega_\mathrm{DM} h^2}{0.12} \simeq & \,  \left( \frac{106.75}{g_*(T_{\mathrm{prod},2})} \right) \left( \frac{r_\mathrm{FO}^2 K_2 (r_\mathrm{FO})}{49.5} \right) \left( \frac{m_\mathrm{DM}}{20\keV} \right).
\end{align}
The relation between the decay width of $P$ and $\xi$ is already discussed in \cref{eq:relation_xi_decay_width}.
Now we have all the ingredients to match between this specific model and our $A_2$--$\xi$ parametrization.
\\
Finally, we want to comment on the potential issue of late time decays of heavy particles which may happen during the epoch of BBN and hence can spoil the abundance of light nuclei by injecting highly energetic particles into 
the thermal plasma \cite{Kawasaki:2017bqm, Hufnagel:2017dgo, Depta:2020zbh}. However, this danger does not appear for our model setup, because we assume that the parent particle decays exclusively into DM via $P \, \to X\,X$. There
is no heating of the SM plasma due to these decays, because the coupling between $X$ and the SM plasma is assumed to be zero. On the contrary, for models which feature decays into SM particles besides DM, $P \to \, X \, S$, these decays tend to be dangerous 
when decaying at temperatures $T\lesssim 1\MeV$. Using \cref{eq:temp_decay_relation,eq:relation_xi_decay_width} this 
can be translated into a bound on $\xi$,
\begin{equation}
 \frac{T}{\mathrm{MeV}} =  \frac{200}{\xi} \frac{m_P}{\mathrm{GeV}}\,.
\end{equation}
For our toy model this bound evaluates to $\xi \leq 2000$ if $P$ decays dominantly into hadrons. 
Instead if leptonic decays dominate, the condition that $T\gtrsim 1\MeV$ can be relaxed substantially, if $A_2\ll 1$ as shown in \cite{Baumholzer:2019twf}. 
Examples for this kind of model setup has been studied in the context of supersymmetric models, where a colored mediator decays slowly to DM after it is frozen-out \cite{Garny:2018ali}. Another example has been studied in Ref.~\cite{Baumholzer:2019twf} using the scotogenic model framework. In this model, light DM is produced by three-body decays of heavy right-handed neutrinos.

\subsection{Toy model III: thermalized + frozen-in parents}
\label{subsec:double_freeze_in}
\begin{figure}[ht!]
	\centering
	\includegraphics[width=0.495\textwidth]{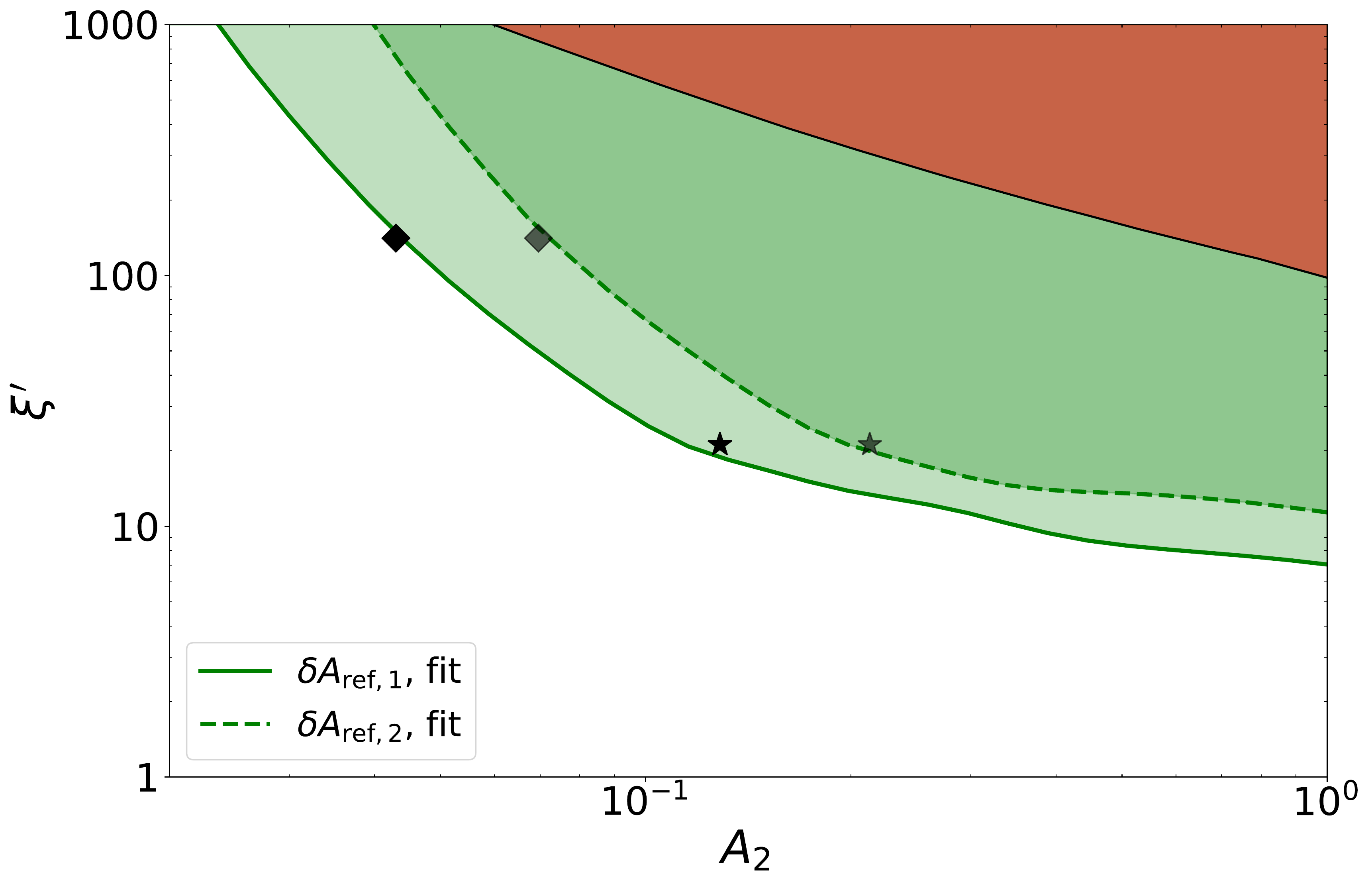}
	\includegraphics[width=0.495\textwidth]{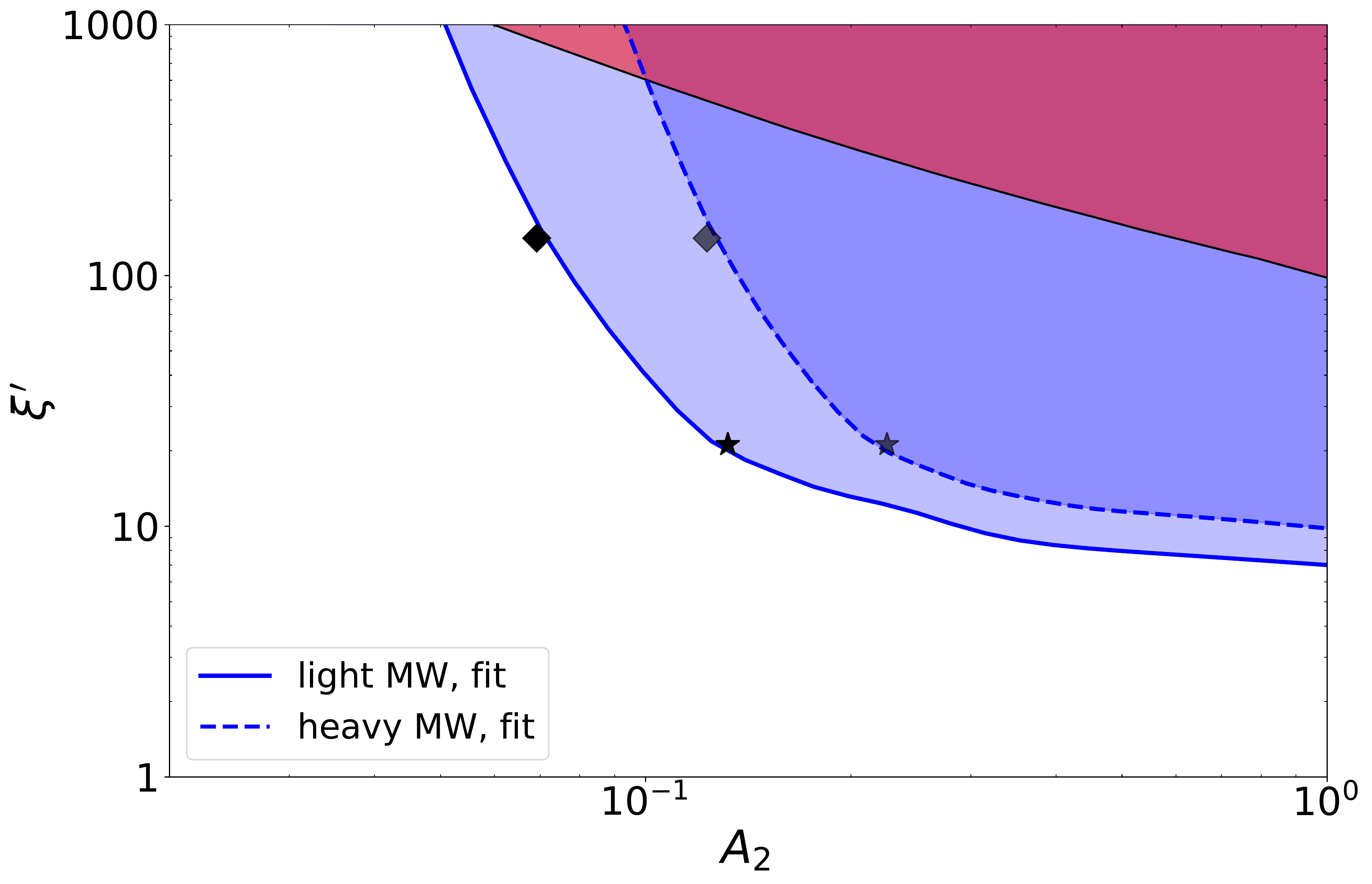}
	\caption{Structure formation limits using the $\delta A_{\mathrm{ref}}$ criterion in the left and a MW subhalo count in the right figure for the third toy model. Everything indicated by the green or blue shaded region is constrained by structure formation and solid lines correspond to strong and dashed lines to weak limits. For comparison a full numerical simulation to extract limits was run for $\xi = 40$ and $300$ and is indicated by the stars and diamonds in both plots.}
	\label{fig:limits_125keV_toy_model}
\end{figure}

\begin{figure}
	\subfloat{\includegraphics[width=0.5\textwidth]{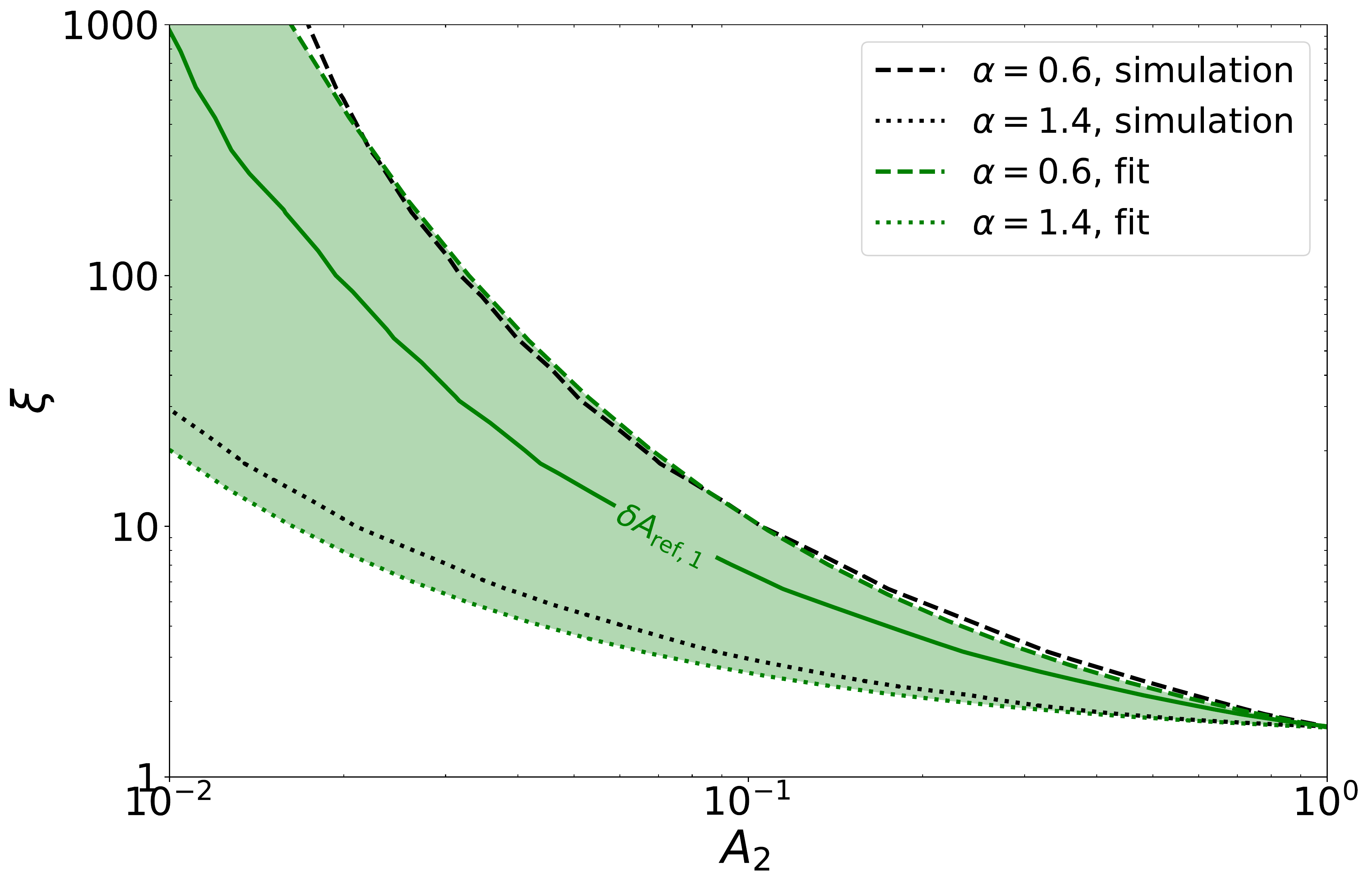}}
	\subfloat{\includegraphics[width=0.5\textwidth]{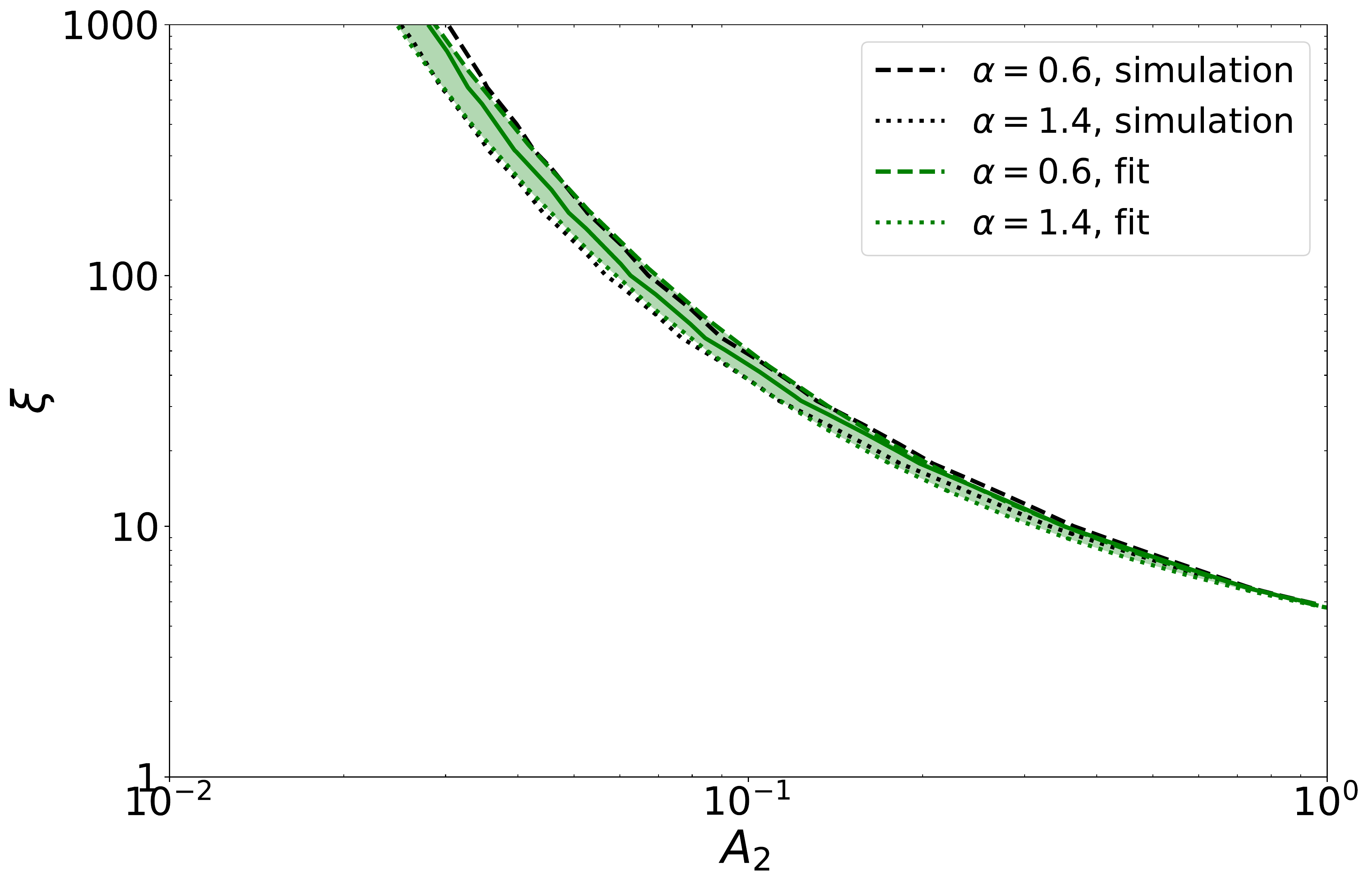}}
	\caption{Effect of changing $\langle z \rangle$ for the $A_1$ set of the DM on the excluded region in the $A_2$--$\xi$ plane based on the $\delta A_{\mathrm{ref},1}$ criterion. The dashed lines correspond to $\alpha = 0.6$, while the dotted curves 
	represents $\alpha = 1.4$. The black curves stem from a numerical analysis and the green ones from the adapted analytical fit as explained in \cref{sec:fit_results}. The left figure is derived for $m_\mathrm{DM}=20\keV$ and the right one for $m_\mathrm{DM}=60\keV$.}
	\label{fig:compare_T1}
\end{figure}

A a last toy model we assume that $m_\mathrm{DM} = 125\keV$, again an amount $A_1$ of DM is produced via thermalized parent decays with $m_S = 1\TeV$, but $A_2$ stems from a frozen-in parent with mass $m_P = 80\GeV$. Similarly to the previous toy model, we have to take a change in $g_*(T)$ into account, although it will impact the limits on $\xi$ at higher values, since $m_P$ is larger in this case
and DM is produced at earlier times compared to the previous case. The results of the fit are shown in \cref{fig:limits_125keV_toy_model} where we use the same color coding as before. Since the DM in this model is heavier 
than in the first toy model, the exclusion limits are shifted to larger $\xi$ values. Further, the bound from $\Delta N_\mathrm{eff}$
is clearly subdominant in this plot due to the rather large DM mass.
\\
A before, we compare our analytical exclusion limits against some benchmark points using a full numerical simulation, but in this case we choose $\xi =40$ and $300$.  Compared to the previous example, the benchmark 
point are closer to the fitted curve here. This is to be expected, as the momentum distribution function $f(x,r)$ (see \cref{eq:momentum_distr_freeze_in}) arising from decays of frozen-in parents only features one distinct peak, because the parent particles are never thermalized and so no early decays are taking place. Hence, 
rescaling $\xi$ by an appropriate factor, as explained around \cref{eq:xi_shifted}, can be safely done even for $A_2 \to 1$.
\\
One can now match these limits onto model parameters describing frozen-in parent particles. Assuming that $P$ couples to a SM particle $Y$ via $\mu \, P \, P \, Y\, Y$, it is produced via an effective coupling, $\displaystyle C_P=M_0/m_P \,\mu^2/64\pi^3$, and the DM yield for this mechanism is given by 
\begin{align}
Y_\mathrm{DM} & = C_P \, \frac{135}{64\pi^2} \, \frac{1}{g_*(T_{\mathrm{prod},2})}\,,\\
A_2 & \simeq \left( \frac{m_\mathrm{DM}}{125\keV} \right) \left( \frac{106.75}{g_*(T_{\mathrm{prod},2})} \right) \left( \frac{C_P}{1.7\cdot 10^{-3}} \right).
\end{align}
\\
When discussing freeze-in scenarios one might want to construct a model where all particles are decoupled from the thermal bath and therefore all of DM is produced by frozen-in parent particles. However, this situation requires some modifications, because the averaged momentum for the first subset generally differs from the case $\langle z \rangle = 2.5$. As pointed out in \cref{sec:fit_results} this different setup can be handled by introducing a shift $\langle z \rangle = 2.5 \, \alpha$. If $\alpha >1$, the DM carries a larger averaged momentum and hence stronger bounds are set on the parameter space. On the other hand, $\alpha < 1$ corresponds to a DM with smaller averaged momenta and weaker structure formation constraints. 
\\
This effect is shown in \cref{fig:compare_T1} where we choose $\alpha = 0.6$ and $1.4$ for two different DM mass choices and compare the corresponding $\delta A_{\mathrm{ref},1}$ exclusion bound to the reference case $\alpha = 1$, but do not change $A_2$. Further, these numerical results shown as the black curves are compared against limits we derived using our analytical fit prescription (green curves), adapted as explained at the end of \cref{sec:model_setup}. As can be seen both approaches agree to a good approximation. Only for rather small DM masses and $\alpha > 1$ both curves differ from each other. However, this is not unexpected, as the structure formation observables are quickly changing in this region of parameter space, because for $m_\mathrm{DM} = 20\keV$ and $\alpha = 1.4$ the absolute mass limit given in \cref{tab:mass_limit} is reached.
Finally, as expected, the impact on the bounds gets weaker if the DM mass is increasing and it becomes negligible for $m_\mathrm{DM} \gtrsim 100\keV$.
\\
A similar effect appears when $g_*(T_{\mathrm{prod},1})$ is smaller compared to our assumption where all SM particles are still in the thermal bath. In that case, one would find a larger temperature $T_1$ compared to the 
photon bath due to smaller reheating effects and accordingly the exclusion curves have to be corrected similar to the case of larger $\langle z \rangle$ values shown in \cref{fig:compare_T1}. For this case, $\alpha$ is defined as the increase in $T_1$, which is given by the ratio $\alpha = (106.75/g_*(T_{\mathrm{prod},1}))^{1/3}$ and for sufficiently late production times the temperature can be twice as large as compared to early decays.\\
While such a scenario has not been discussed widely in the literature, it can occur for example in the model of~\cite{Merle_2015,Konig:2016dzg}, where right-handed neutrino DM is produced by decays of a heavy scalar particle. Depending on the respective couplings to the SM, these scalars do not necessarily thermalize before the onset of DM freeze-in.

\section{Summary}
\label{sec:summary}

Many extensions of the SM introduce entire dark sectors with several new particles and interactions among them. 
Therefore, it is natural to ask the question what happens if the DM is produced via 
different production channels, leading to DM composed of several subsets, each characterized by its own temperature. Depending on the size and the nature of the involved couplings and particles, these decays can easily take place at late times in the early Universe and 
give rise to an increased DM temperature. We examined such a scenario as model-independent as possible to allow for an easy comparison with specific warm DM models. For this purpose, we assumed that the DM is produced
by two different decay channels. One is due to decays of thermalized parents at rather early times, while the second contribution stems from decays happening at later times.
Our setup (referred to as 2TDM) is parameterized by two key parameters: the abundance $A_2$ of the second subset produced at later times and the temperature ratio between both DM subsets, denoted by $\xi$.
\\
The impact of such a model setup on the formation of structures in the universe was evaluated. Specifically we derived predictions for the number of MW subhalos and the flux power spectrum and compared them against observations.
Based on these, limits on the parameters $\xi$ and $A_2$ were derived for DM masses between $20$--$500\keV$. For $\xi = 1$, i.e. a single DM temperature setup, our limits on the DM mass are up to $11\keV$ using Lyman-$\alpha$ measurements
and up to $13\keV$ counting the observed MW subhalos and using the MW mass derived from recent GAIA measurements. Typically, we could probe and constrain parameters for $A_2$ between $1$ and $0.01$ for temperature ratios up to $1000$
and in general, DM with a high temperature $T_2$ can only make up a few percent of the total DM number density.
\\
We presented an analytical fit for the respective exclusion limit and discussed further steps how to extract limits on 
specific model realizations. One focus was the incorporation of a change in the number of entropic degrees of freedom during the time of production of the DM species, as this impacts the interpretation of the fitted results.
As an example we considered different examples and compared our analytical prediction against numerical results, where we made direct use of appropriate momentum distribution functions. Our procedure
showed a good agreement between the analytical fit and actual results. Hence, it allowed us to predict limits on the temperature of a warmer DM fraction and its abundance without extensive simulations.\\
Lastly, we commented on the treatment of DM production via three-body decays inside our framework and how one can calculate exclusion limits for these cases applying rescaled results.

\section{Acknowledgments}
The research of SB and PS is supported by the Cluster of Excellence Precision Physics, Fundamental Interactions, 
and Structure of Matter(PRISMA+ EXC 2118/1) funded by the German Research Foundation(DFG) within the German Excellence Strategy (Project ID 39083149),  
and by grant 05H18UMCA1 of the German Federal Ministry for Education and Research (BMBF).

\section{Appendix}

\subsection{Momentum distribution function of out of equilibrium parents}
\label{subsec:decay_freeze_out_scalar}
If the parent particle is sufficiently long-lived and has a sizable coupling $C_P$, it thermalizes and its decay will happen after it drops out of the thermal bath. After the time of freeze-out, $r_\text{FO}$, the momentum distribution function of the parent is given by
\begin{align}
f_P(x,r) = & \, f_\text{eq}(x,r),\quad r<r_\text{FO}\,,\\
f_P(x,r) = & \, f_\text{eq}(x,r_\text{FO}) \left( \frac{r + \sqrt{r^2 + x^2}}{r_\text{FO} + \sqrt{r_\text{FO}^2 + x^2 }} \right)^{C_\Gamma x^2/2}\times \\ 
\times & \, e^{-C_\Gamma(r\sqrt{r^2 + x^2} - r_\text{FO}\sqrt{r_\text{FO}^2 + x^2 })/2}, \quad r>r_\text{FO}\,.
\label{eq:momentum_distr_out_of_equi}
\end{align}
\subsection{Momentum distribution function of never thermalized parents}
\label{subsec:decay_non_thermal_scalar}
Weakly coupled parent particles with $C_P\ll 1$ never reach thermal equilibrium, but rather freeze-in before they start to decay. Their momentum distribution can be derived as
\begin{equation}
f_P(r,x) = C_P \int\limits_0^r \dd \rho \, \rho K_1(\rho) \frac{\text{exp}(-\sqrt{\rho^2 + x^2})}{\sqrt{\rho^2 + x^2}} \left( \frac{e^{\rho \sqrt{\rho^2 + x^2}}}{e^{r \sqrt{r^2 + x^2}}}  \left( \frac{\rho + \sqrt{\rho^2 + x^2}}{r + \sqrt{r^2 + x^2}} \right)^{x^2/2} \right)^{C_\Gamma/2}.
\label{eq:momentum_distr_freeze_in}
\end{equation}

\subsection{Change in $\langle z \rangle$ due to $g_*(T)$}

\begin{figure}[ht]
 \centering
 \includegraphics[width=0.95\textwidth]{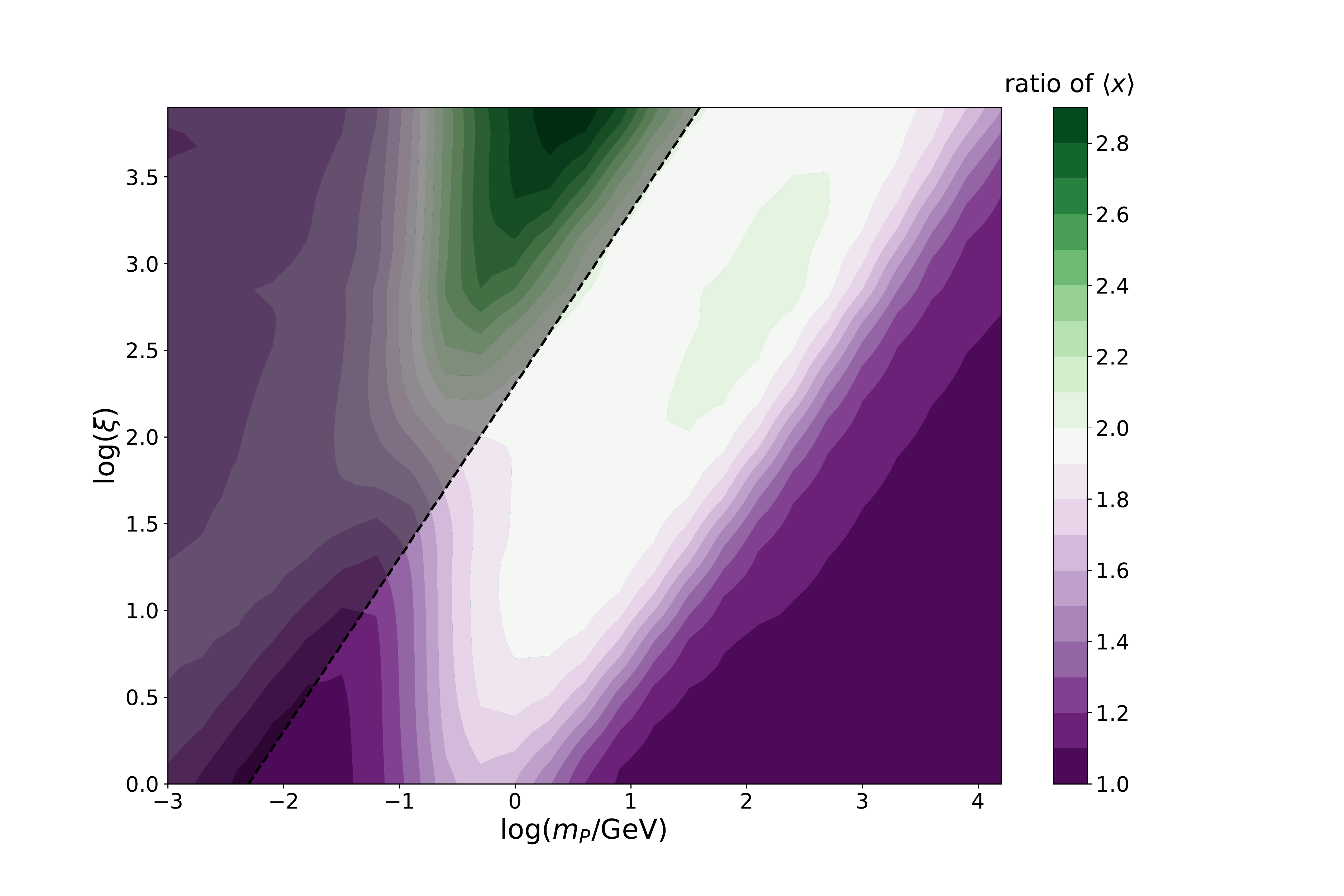}
 \caption{Ratio of $\langle x \rangle$ derived comparing \cref{eq:boltzman_varying_dof} against \cref{eq:master_equation}, i.e. comparing a time-dependent $g_*(T)$ against constant case $g_*(T)=106.75$. As can be seen, the averaged value is shifted to larger values, especially for $m_P \sim 1\GeV$ and large values of $\xi$, because in that
 region $g_*(T)$ is changing rapidly. However, the effect becomes less prominent for heavier or lighter parent particles. It can also be seen that $\langle x \rangle$ stays constant if both, $m_P$ and $\xi$ are increased. 
 The black shaded region indicates decays which would take place after BBN.}
 \label{fig:pt_ratio}
\end{figure}

As explained in \cref{sec:modifications_variable_dof}, the time-dependence of the entropic degrees of freedom, $g_*(T)$, generally shifts the DM momentum distribution function towards larger momenta. To examine this effect for the case of out of equilibrium decays, the average value $\langle z \rangle$ is calculated, taking a change in $g_*(T)$ into account, and compared against $\langle x \rangle$, where 
$g_*(T)$ are kept fixed in the calculation. The result is shown in \cref{fig:pt_ratio}, where we varied the mass of the parent particle between $1\MeV$ and roughly $10\TeV$ for $\xi$ between $1$ and $6000$. As can be seen, 
at rather small $\xi$ the effect of a change in $g_*(T)$ starts to be dominant at $m_P<10\GeV$; below this mass the averaged momentum can be even twice as large as compared to the case where $g_*(T)$ are treated as constant.
However, increasing $\xi$ will give rise to a shift in $\langle z \rangle$ even for rather large parent particle masses. This behavior is expected, because if DM is produced at sufficiently early times by decays of very heavy parent
particles, $g_*(T)$ stays approximately constant. However, demanding that this DM should have a large $\xi$ as well, requires that it is produced at later times in the temperature and the effect of a change in $g_*(T)$ becomes relevant.

\clearpage
\subsection{Details on fit parameters}
\label{app:details_fit_pi}
As explained in \cref{sec:fit_results} we fit the exclusion contours using \cref{eq:fit_equation} and derive the fit parameters $p_i$ for $m_\mathrm{DM}$ between $20$--$500\keV$. The mass-dependent $p_i$ are than fitted using \cref{eq:poly_fit}.
\\
In total we are using eight fit parameters in \cref{eq:fit_equation} for our final fit and the results are summarized in \cref{tab:fit_results_pi}.


\begin{table}[ht!]
	\centering
	\begin{tabu}{|c|c|c|c|c|}
		\hline
		&	$\delta A_{\mathrm{ref},1}$		&	$\delta A_{\mathrm{ref},2}$		&	\textit{light} MW mass	&	\textit{heavy} MW mass		\\	
		\hline
		$a_0$	&	$1.23 $					&	$1.15 $				&		$0.431 $				&	$0.494$	\\
		$b_0$	&	$-15.3 $				&	$-11.6 $				&		$-5.382 $				&	$-5.35 $			\\
		$c_0$	&	$0.253 \cdot 10^{-4}$	& 	$-7.47 \cdot 10^{-4}$	&		$-8.96 \cdot 10^{-4}$	&	$-11.9  \cdot 10^{-4}$ \\
		$d_0$	&	$0.815 \cdot 10^{-6}$	&	$1.75  \cdot 10^{-6}$	&		$1.24 \cdot 10^{-6}$	&	$1.37  \cdot 10^{-6}$	\\
		\hline
		$a_1$	&	$0.495$					&	$0.557$					&		$0.903$					&	$1.029 $				\\
		$b_1$	&	$1.68 $					&	$2.01$						&		$3.670 $				&	$4.47 $				\\
		$c_1$	&	$-1.07  \cdot 10^{-4}$	&	$1.52  \cdot 10^{-4}$	&		$9.39  \cdot 10^{-4}$	&	$15.6 \cdot 10^{-4}$	\\
		$d_1$	&	$-0.522  \cdot 10^{-7}$	&	$-4.48 \cdot 10^{-7}$	&		$-12.8 \cdot 10^{-7}$	&	$-14.7 \cdot 10^{-7}$	\\
		\hline
	\end{tabu}
	\caption{Final fit parameters for all four exclusion contours, according to \cref{eq:poly_fit}. The DM ranges between $20$--$500\keV$.}
	\label{tab:fit_results_pi}
\end{table}


\clearpage
\subsection{Flow chart}
\begin{figure}[ht!]
	\centering
	\includegraphics[width=0.5\textwidth]{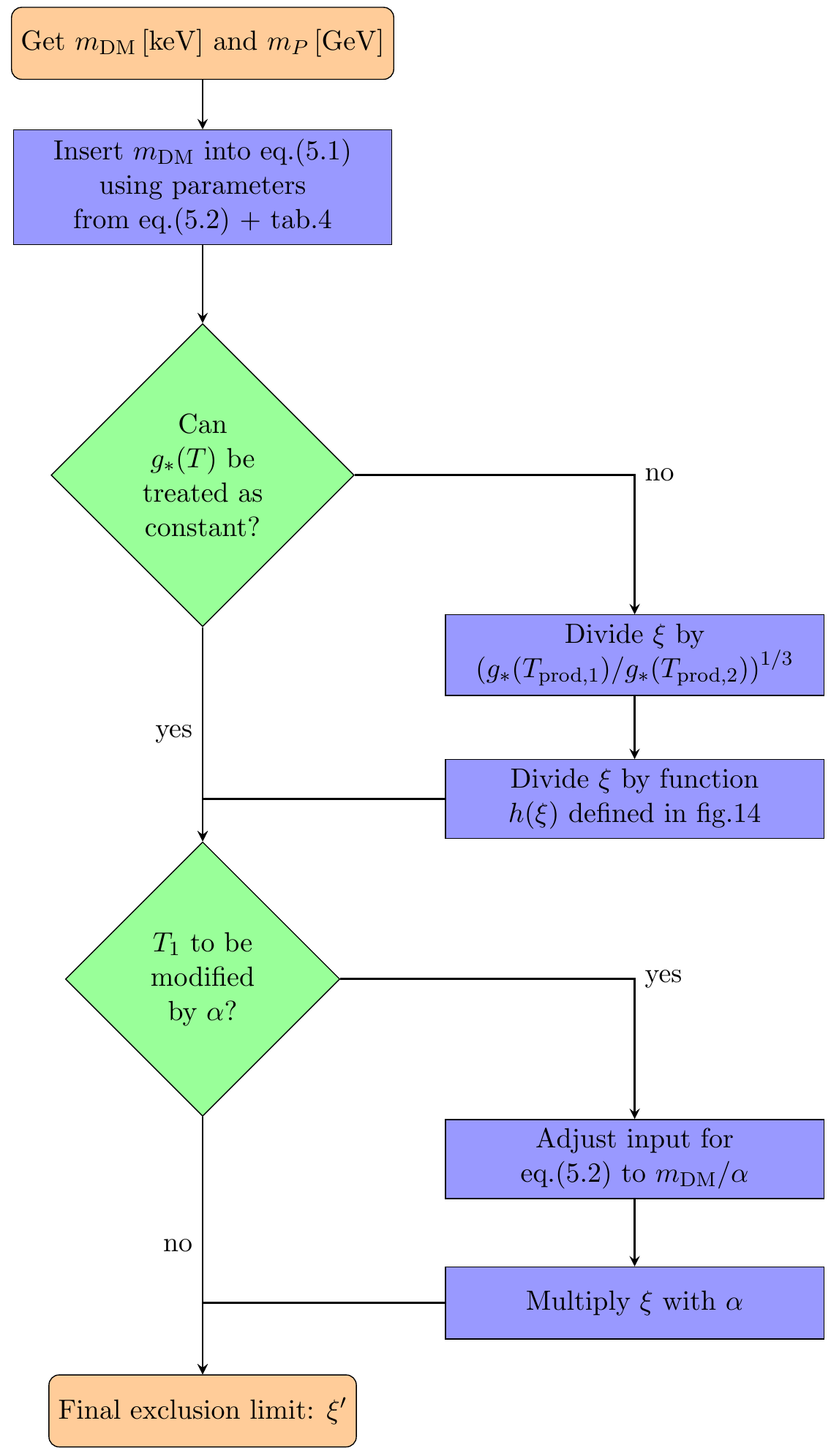}
	\caption{Flow chart of the guide line on how to properly modify limits on $\xi$ when the assumption that $g_*(T) = 106.75=\mathrm{const}$
		is not valid during the production of $A_2$, or the first DM subset features a different temperature $T_1$ either due to smaller $g_*(T)$ during its production or a slightly
		different production mechanism. More details on the procedure are given in \cref{sec:fit_results}.}
	\label{fig:flow_chart}
\end{figure}

\clearpage
\bibliography{Draft_ref}
\bibliographystyle{JHEP}
\end{document}